\pdfoutput=1
\documentclass[12pt,a4paper]{article}
\usepackage{ifthen} % for conditional statements
\newboolean{pdflatex}
\setboolean{pdflatex}{true} % False for eps figures 

\newboolean{articletitles}
\setboolean{articletitles}{true} % False removes titles in references

\newboolean{uprightparticles}
\setboolean{uprightparticles}{false} %True for upright particle symbols

\newboolean{inbibliography}
\setboolean{inbibliography}{false} %True once you enter the bibliography

\usepackage[top=1in, bottom=1.25in, left=1in, right=1in]{geometry}

\usepackage{microtype}
\usepackage{lineno}  % for line numbering during review
\usepackage{xspace} % To avoid problems with missing or double spaces after
\usepackage{caption} %these three command get the figure and table captions automatically small

\usepackage{graphicx}  % to include figures (can also use other packages)
\usepackage{color}
\usepackage{colortbl}
\graphicspath{{./figs/}} % Make Latex search fig subdir for figures

\usepackage{amsmath} % Adds a large collection of math symbols
\usepackage{amssymb}
\usepackage{amsfonts}
\usepackage{upgreek} % Adds in support for greek letters in roman typeset

\newcommand*\patchAmsMathEnvironmentForLineno[1]{%
\expandafter\let\csname old#1\expandafter\endcsname\csname #1\endcsname
\expandafter\let\csname oldend#1\expandafter\endcsname\csname
end#1\endcsname
 \renewenvironment{#1}%
   {\linenomath\csname old#1\endcsname}%
   {\csname oldend#1\endcsname\endlinenomath}%
}
\newcommand*\patchBothAmsMathEnvironmentsForLineno[1]{%
  \patchAmsMathEnvironmentForLineno{#1}%
  \patchAmsMathEnvironmentForLineno{#1*}%
}
\AtBeginDocument{%
\patchBothAmsMathEnvironmentsForLineno{equation}%
\patchBothAmsMathEnvironmentsForLineno{align}%
\patchBothAmsMathEnvironmentsForLineno{flalign}%
\patchBothAmsMathEnvironmentsForLineno{alignat}%
\patchBothAmsMathEnvironmentsForLineno{gather}%
\patchBothAmsMathEnvironmentsForLineno{multline}%
\patchBothAmsMathEnvironmentsForLineno{eqnarray}%
}

\usepackage{hyperref}    % Hyperlinks in references
\usepackage[all]{hypcap} % Internal hyperlinks to floats.

\usepackage{xspace} 
\usepackage{upgreek}

\def\lhcb {\mbox{LHCb}\xspace}

\def\MagUp {\mbox{\em Mag\kern -0.05em Up}\xspace}

\ifthenelse{\boolean{uprightparticles}}%
{

 \def\Pmu         {\ensuremath{\upmu}\xspace}

 \def\Ppi         {\ensuremath{\uppi}\xspace}

 \def\Ppsi        {\ensuremath{\uppsi}\xspace}

 \def\PDelta      {\ensuremath{\Delta}\xspace}                 
 \def\PXi      {\ensuremath{\Xi}\xspace}                 
 \def\PLambda      {\ensuremath{\Lambda}\xspace}                 
 \def\PSigma      {\ensuremath{\Sigma}\xspace}                 
 \def\POmega      {\ensuremath{\Omega}\xspace}                 
 \def\PUpsilon      {\ensuremath{\Upsilon}\xspace}                 
 
 \def\PB      {\ensuremath{\mathrm{B}}\xspace}                 
                  
 \def\PD      {\ensuremath{\mathrm{D}}\xspace}

 \def\PJ      {\ensuremath{\mathrm{J}}\xspace}                 
 \def\PK      {\ensuremath{\mathrm{K}}\xspace}

 \def\Pb      {\ensuremath{\mathrm{b}}\xspace}                 
 \def\Pc      {\ensuremath{\mathrm{c}}\xspace}

 \def\Pi      {\ensuremath{\mathrm{i}}\xspace}

 \def\Ps      {\ensuremath{\mathrm{s}}\xspace}

}
{

 \def\Pmu         {\ensuremath{\mu}\xspace}

 \def\Ppi         {\ensuremath{\pi}\xspace}

 \def\Ppsi        {\ensuremath{\psi}\xspace}                 
                  
 \mathchardef\PDelta="7101
 \mathchardef\PXi="7104
 \mathchardef\PLambda="7103
 \mathchardef\PSigma="7106
 \mathchardef\POmega="710A
 \mathchardef\PUpsilon="7107
                  
 \def\PB      {\ensuremath{B}\xspace}                 
                  
 \def\PD      {\ensuremath{D}\xspace}

 \def\PJ      {\ensuremath{J}\xspace}                 
 \def\PK      {\ensuremath{K}\xspace}

 \def\Pb      {\ensuremath{b}\xspace}                 
 \def\Pc      {\ensuremath{c}\xspace}

 \def\Pi      {\ensuremath{i}\xspace}

 \def\Ps      {\ensuremath{s}\xspace}

}

\makeatletter
\ifcase \@ptsize \relax% 10pt
  \newcommand{\miniscule}{\@setfontsize\miniscule{4}{5}}% \tiny: 5/6
\or% 11pt
  \newcommand{\miniscule}{\@setfontsize\miniscule{5}{6}}% \tiny: 6/7
\or% 12pt
  \newcommand{\miniscule}{\@setfontsize\miniscule{5}{6}}% \tiny: 6/7
\fi
\makeatother

\DeclareRobustCommand{\optbar}[1]{\shortstack{{\miniscule (\rule[.5ex]{1.25em}{.18mm})}
  \\ [-.7ex] $#1$}}

\def\mup        {{\ensuremath{\Pmu^+}}\xspace}
\def\mun        {{\ensuremath{\Pmu^-}}\xspace} % muon negative (\mum is taken)
\def\mumu       {{\ensuremath{\Pmu^+\Pmu^-}}\xspace}

\def\squark    {{\ensuremath{\Ps}}\xspace}

\def\cquark    {{\ensuremath{\Pc}}\xspace}
\def\cquarkbar {{\ensuremath{\overline \cquark}}\xspace}
\def\ccbar     {{\ensuremath{\cquark\cquarkbar}}\xspace}
\def\bquark    {{\ensuremath{\Pb}}\xspace}
\def\bquarkbar {{\ensuremath{\overline \bquark}}\xspace}

\def\pion   {{\ensuremath{\Ppi}}\xspace}

\def\pip    {{\ensuremath{\pion^+}}\xspace}
\def\pim    {{\ensuremath{\pion^-}}\xspace}

\def\kaon    {{\ensuremath{\PK}}\xspace}
  \def\Kbar    {{\kern 0.2em\overline{\kern -0.2em \PK}{}}\xspace}

\def\KorKbar    {\kern 0.18em\optbar{\kern -0.18em K}{}\xspace}

\def\Kp      {{\ensuremath{\kaon^+}}\xspace}
\def\Km      {{\ensuremath{\kaon^-}}\xspace}

\def\Kstarz  {{\ensuremath{\kaon^{*0}}}\xspace}
\def\Kstarzb {{\ensuremath{\Kbar{}^{*0}}}\xspace}
\def\Kstar   {{\ensuremath{\kaon^*}}\xspace}

  \def\Dbar    {{\kern 0.2em\overline{\kern -0.2em \PD}{}}\xspace}

\def\DorDbar    {\kern 0.18em\optbar{\kern -0.18em D}{}\xspace}

\def\B       {{\ensuremath{\PB}}\xspace}
\def\Bbar    {{\ensuremath{\kern 0.18em\overline{\kern -0.18em \PB}{}}}\xspace}

\def\BorBbar    {\kern 0.18em\optbar{\kern -0.18em B}{}\xspace}
\def\Bz      {{\ensuremath{\B^0}}\xspace}
\def\Bzb     {{\ensuremath{\Bbar{}^0}}\xspace}

\def\Bd      {{\ensuremath{\B^0}}\xspace}
\def\Bs      {{\ensuremath{\B^0_\squark}}\xspace}

\def\jpsi     {{\ensuremath{{\PJ\mskip -3mu/\mskip -2mu\Ppsi\mskip 2mu}}}\xspace}

  \def\Y#1S{\ensuremath{\PUpsilon{(#1S)}}\xspace}% no space before {...}!

\def\Lbar        {{\ensuremath{\kern 0.1em\overline{\kern -0.1em\PLambda}}}\xspace}
\def\LorLbar    {\kern 0.18em\optbar{\kern -0.18em \PLambda}{}\xspace}

\def\BF         {{\ensuremath{\mathcal{B}}}\xspace}

\newcommand{\decay}[2]{\ensuremath{#1\!\to #2}\xspace}         % {\Pa}{\Pb \Pc}

\def\to                 {\ensuremath{\rightarrow}\xspace}

\def\qsq       {{\ensuremath{q^2}}\xspace}

\def\CP                {{\ensuremath{C\!P}}\xspace}

\def\BdToJPsiKst  {\decay{\Bd}{\jpsi\Kstarz}}
\def\BdToJPsiKst  {\decay{\Bd}{\jpsi\Kstarz}}

\def\AT#1     {\ensuremath{A_{\mathrm{T}}^{#1}}\xspace}           % 2

\def\ctl       {\ensuremath{\cos{\theta_\ell}}\xspace}
\def\ctk       {\ensuremath{\cos{\theta_K}}\xspace}

\def\C#1      {\ensuremath{\mathcal{C}_{#1}}\xspace}                       % 9
\def\Cp#1     {\ensuremath{\mathcal{C}_{#1}^{'}}\xspace}                    % 7
\def\Ceff#1   {\ensuremath{\mathcal{C}_{#1}^{\mathrm{(eff)}}}\xspace}        % 9  
\def\Cpeff#1  {\ensuremath{\mathcal{C}_{#1}^{'\mathrm{(eff)}}}\xspace}       % 7
\def\Ope#1    {\ensuremath{\mathcal{O}_{#1}}\xspace}                       % 2
\def\Opep#1   {\ensuremath{\mathcal{O}_{#1}^{'}}\xspace}                    % 7

\newcommand{\tev}{\ifthenelse{\boolean{inbibliography}}{\ensuremath{~T\kern -0.05em eV}\xspace}{\ensuremath{\mathrm{\,Te\kern -0.1em V}}}\xspace}
\newcommand{\gev}{\ensuremath{\mathrm{\,Ge\kern -0.1em V}}\xspace}
\newcommand{\mev}{\ensuremath{\mathrm{\,Me\kern -0.1em V}}\xspace}
\newcommand{\kev}{\ensuremath{\mathrm{\,ke\kern -0.1em V}}\xspace}
\newcommand{\ev}{\ensuremath{\mathrm{\,e\kern -0.1em V}}\xspace}
\newcommand{\gevc}{\ensuremath{{\mathrm{\,Ge\kern -0.1em V\!/}c}}\xspace}
\newcommand{\mevc}{\ensuremath{{\mathrm{\,Me\kern -0.1em V\!/}c}}\xspace}
\newcommand{\gevcc}{\ensuremath{{\mathrm{\,Ge\kern -0.1em V\!/}c^2}}\xspace}
\newcommand{\gevgevcccc}{\ensuremath{{\mathrm{\,Ge\kern -0.1em V^2\!/}c^4}}\xspace}
\newcommand{\mevcc}{\ensuremath{{\mathrm{\,Me\kern -0.1em V\!/}c^2}}\xspace}

\def\mum  {\ensuremath{{\,\upmu\mathrm{m}}}\xspace}

\def\invfb   {\ensuremath{\mbox{\,fb}^{-1}}\xspace}

\def\deriv {\ensuremath{\mathrm{d}}}

\def\gsim{{~\raise.15em\hbox{$>$}\kern-.85em
          \lower.35em\hbox{$\sim$}~}\xspace}
\def\lsim{{~\raise.15em\hbox{$<$}\kern-.85em
          \lower.35em\hbox{$\sim$}~}\xspace}

\def\sPlot{\mbox{\em sPlot}\xspace}
\def\pt         {\mbox{$p_{\mathrm{ T}}$}\xspace}

\def\evtgen     {\mbox{\textsc{EvtGen}}\xspace}

\def\geant      {\mbox{\textsc{Geant4}}\xspace}

\def\photos     {\mbox{\textsc{Photos}}\xspace}

\def\pythia     {\mbox{\textsc{Pythia}}\xspace}

\def\tell1  {TELL1\xspace}
\def\ukl1   {UKL1\xspace}

\usepackage{cite} % Allows for ranges in citations
\usepackage{mciteplus}

\usepackage{longtable} % only for template; not usually to be used in PAPERs

\usepackage{amsmath}
\usepackage{subfigure}
\def\mkpi  {\ensuremath{m(\Kp\pim)}\xspace}
\def\mkpimm{\ensuremath{m(\Kp\pim\mup\mun)}\xspace}
\def\kpi{\ensuremath{\Kp\pim}\xspace}
\def\kpimm{\ensuremath{\Kp\pim\mup\mun}\xspace}
\def\mkpip  {\ensuremath{m'(\Kp\pim)}\xspace}
\def\qsqp  {\ensuremath{\qsq'}\xspace}
\def\phip  {\ensuremath{\phi'}\xspace}

\def\BdToKpimm    {\decay{\Bd}{\Kp\pim\mumu}}

\def\BsToJPsiKst  {\decay{\Bs}{\jpsi\Kstarzb}}

\def\BdToJPsiKpi  {\decay{\Bd}{\jpsi \Kp\pim}}

\def\BdToJPsiKstP  {\decay{\Bd}{\jpsi\kaon^{*}(892)^{0}}}
\def\KstP  {\ensuremath{\kaon^{*}(892)^{0}}\xspace}

\def\thetal {\ensuremath{\theta_\ell}\xspace}
\def\thetak {\ensuremath{\theta_K}\xspace}

\newcommand {\img} {\, Im}
\newcommand {\rel} {\, Re}

\def\hzsq        {\ensuremath{|H^L_0|^2}\xspace}

\def\ssq         {\ensuremath{|S^L|^2}\xspace}
\def\dzsq        {\ensuremath{|D^L_0|^2}\xspace}

\def\hpasq        {\ensuremath{|H^L_\parallel|^2}\xspace}
\def\hpesq        {\ensuremath{|H^L_\perp|^2}\xspace}
\def\dpasq        {\ensuremath{|D^L_\parallel|^2}\xspace}
\def\dpesq        {\ensuremath{|D^L_\perp|^2}\xspace}

\def\rhzdz       {\ensuremath{Re(H^L_0 D^{L\ast}_0)}\xspace}
\def\rshz        {\ensuremath{Re(S^L H_0^{L \ast})}\xspace}

\def\rsdz        {\ensuremath{Re(S^L D_0^{L \ast})}\xspace}

\begin{document}

%%%%%%%%%%%%%%%%%%%%%%%%%
%%%%% Title     %%%%%%%%%
%%%%%%%%%%%%%%%%%%%%%%%%%
\renewcommand{\thefootnote}{\fnsymbol{footnote}}
\setcounter{footnote}{1}

% %%%%%%% CHOOSE TITLE PAGE--------
%\onecolumn
%\input{title-LHCb-INT}
%\input{title-LHCb-ANA}
%\input{title-LHCb-CONF}
% $Id: title-LHCb-PAPER.tex 85745 2016-01-05 10:00:22Z lafferty $
% ===============================================================================
% Purpose: LHCb-PAPER journal paper title page template
% Author: 
% Created on: 2010-09-25
% ===============================================================================

%%%%%%%%%%%%%%%%%%%%%%%%%
%%%%%  TITLE PAGE  %%%%%%
%%%%%%%%%%%%%%%%%%%%%%%%%
\begin{titlepage}
\pagenumbering{roman}

% Header ---------------------------------------------------
\vspace*{-1.5cm}
\centerline{\large EUROPEAN ORGANIZATION FOR NUCLEAR RESEARCH (CERN)}
\vspace*{1.5cm}
\noindent
\begin{tabular*}{\linewidth}{lc@{\extracolsep{\fill}}r@{\extracolsep{0pt}}}
\ifthenelse{\boolean{pdflatex}}% Logo format choice
{\vspace*{-2.7cm}\mbox{\!\!\!\includegraphics[width=.14\textwidth]{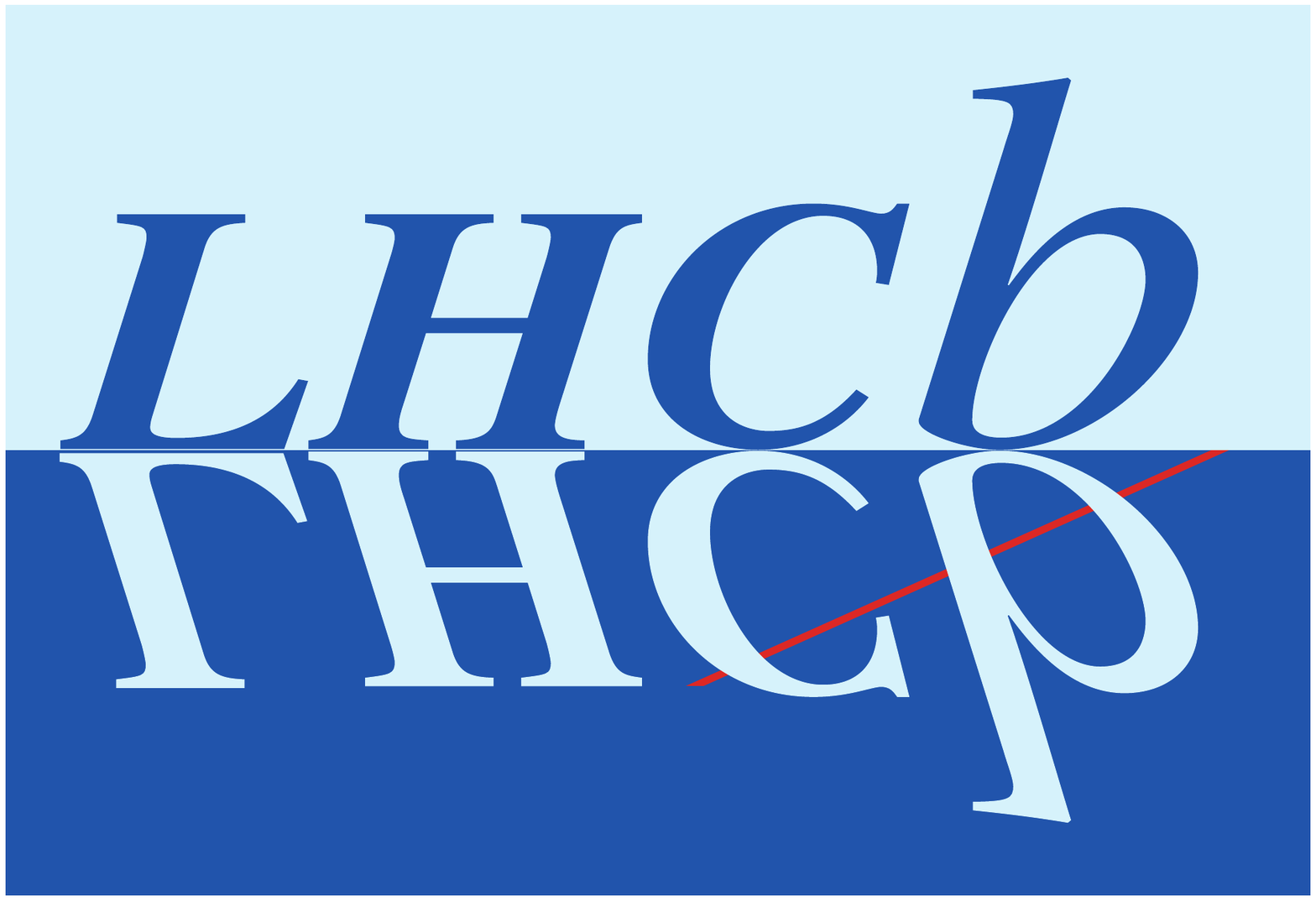}} & &}%
{\vspace*{-1.2cm}\mbox{\!\!\!\includegraphics[width=.12\textwidth]{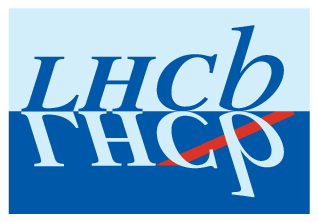}} & &}%
\\
 & & CERN-EP-2016-215 \\  % ID 
 & & LHCb-PAPER-2016-025 \\  % ID 
 & & December 7, 2016 \\ % Date - Can also hardwire e.g.: 23 March 2010
 %& & \today \\ % Date - Can also hardwire e.g.: 23 March 2010
 & & \\
% not in paper \hline
\end{tabular*}

\vspace*{1.0cm}

% Title --------------------------------------------------
{\normalfont\bfseries\boldmath\huge
\begin{center}
  Differential branching fraction and angular moments analysis of the decay $\B^0 \!\to  K^{+}\pi^{-}\mu^{+}\mu^{-}$ in the $K^{*}_{0,2}(1430)^{0}$ region
\end{center}
}

\vspace*{1.0cm}

% Authors -------------------------------------------------
\begin{center}
%In the footnote, replace 'paper' by 'letter' in case of submission to PRL or PLB 
The LHCb collaboration\footnote{Authors are listed at the end of this paper.}
\end{center}

\vspace{\fill}

% Abstract -----------------------------------------------
\begin{abstract}
  \noindent Measurements of the differential branching fraction and angular moments of the decay $\B^0 \!\to  K^{+}\pi^{-}\mu^{+}\mu^{-}$ in the $K^{+}\pi^{-}$ invariant mass range \mbox{$1330<m(\Kp\pim)<1530~\mathrm{\,Me\kern -0.1em V\!/}c^{2}$} are presented. Proton-proton collision data are used, corresponding to an integrated luminosity of 3$\mbox{\,fb}^{-1}$ collected by the LHCb experiment. Differential branching fraction measurements are reported in five bins of the invariant mass squared of the dimuon system, $q^{2}$, between $0.1$ and $8.0{\mathrm{\,Ge\kern -0.1em V^2\!/}c^4}$. For the first time, an angular analysis sensitive to the S-, P- and D-wave contributions of this rare decay is performed. The set of 40 normalised angular moments describing the decay is presented for the $q^{2}$ range $1.1$--$6.0{\mathrm{\,Ge\kern -0.1em V^2\!/}c^4}$. 

\end{abstract}

\vspace*{2.0cm}

\begin{center}
  Published in JHEP {\bf 12} (2016) 065 
  %Submitted to JHEP% / Phys.~Rev.~D / Phys.~Rev.~Lett. / Phys.~Lett.~B / Eur.~Phys.~J.~C / Nucl.~Phys.~B 
\end{center}

\vspace{\fill}

{\footnotesize 
\centerline{\copyright~CERN on behalf of the \lhcb collaboration, licence \href{http://creativecommons.org/licenses/by/4.0/}{CC-BY-4.0}.}}
\vspace*{2mm}

\end{titlepage}

%%%%%%%%%%%%%%%%%%%%%%%%%%%%%%%%
%%%%%  EOD OF TITLE PAGE  %%%%%%
%%%%%%%%%%%%%%%%%%%%%%%%%%%%%%%%

%  empty page follows the title page ----
\newpage
\setcounter{page}{2}
\mbox{~}
%\newpage
%
%% Author List ----------------------------
%%  You need to get a new author list!
%\input{LHCb_authorlist.tex}
%
%The author list for journal publications is provided by the Membership Committee shortly after 'approval to go to paper' has been given.
%%It will be made available on the page
%%\verb!http://www.physik.uzh.ch/~strauman/forMemCo/LHCb-PAPER-XXXX-XXX/! .
%It will be sent to you by email shortly after a paper number has beens assigned.
%The author list should be included already at first circulation, 
%to allow new members of the collaboration to verify whether they have been included correctly.
%Occasionally a misspelled name is corrected or associated institutions become full members.
%In that case, a new author list will be sent to you.
%In case line numbering doesn't work well after including the authorlist, try moving the \verb!\bigskip! after the last author to a separate line.
%
%
%The authorship for Conference Reports should be ``The LHCb
%  collaboration'', with a footnote giving the name(s) of the contact
%  author(s), but without the full list of collaboration names.

\cleardoublepage

%\twocolumn
% %%%%%%%%%%%%% ---------

\renewcommand{\thefootnote}{\arabic{footnote}}
\setcounter{footnote}{0}

%%%%%%%%%%%%%%%%%%%%%%%%%%%%%%%%
%%%%%  Table of Content   %%%%%%
%%%%%%%%%%%%%%%%%%%%%%%%%%%%%%%%
%%%% Uncomment next 2 lines if desired
%\tableofcontents
%\cleardoublepage

%%%%%%%%%%%%%%%%%%%%%%%%%
%%%%% Main text %%%%%%%%%
%%%%%%%%%%%%%%%%%%%%%%%%%

\pagestyle{plain} % restore page numbers for the main text
\setcounter{page}{1}
\pagenumbering{arabic}

%% Uncomment during review phase. 
%% Comment before a final submission.
%\linenumbers

% You can include short sections directly in the main tex file.
% However, for larger papers it is desirable to split the text into
% several semiautonomous files, which can be revised independently.
% This is especially useful when developing a document in
% collaboration with several people, since then different parts can be
% edited independently.  This type of file organization is shown here.
% 

\section{Introduction}
\label{sec:introduction}

The decay \BdToKpimm is a flavour-changing neutral-current process.\footnote{The inclusion of charge conjugate processes is implied, unless otherwise noted.} In the Standard Model (SM), the leading order transition amplitudes are described by electroweak penguin or box diagrams.  In extensions to the SM, new heavy particles can contribute to loop diagrams and modify observables such as branching fractions and angular distributions.

The previous angular analyses of \BdToKpimm performed by the \lhcb collaboration~\cite{LHCb-PAPER-2011-020,LHCb-PAPER-2013-019,LHCb-PAPER-2013-037,LHCb-PAPER-2015-051} focused on the $K^{+}\pi^{-}$ invariant mass range $796<\mkpi<996\mevcc$ where the decay proceeds predominantly via the P-wave process $\decay{\Kstar(892)^{0}}{\kpi}$. 
A global analysis of the \CP-averaged angular observables measured in the \lhcb Run 1 data sample indicated differences from SM predictions at the level of 3.4 standard deviations~\cite{LHCb-PAPER-2015-051}. 
This measurement is widely discussed in the literature (see, for instance, \cite{DescotesGenon:2013wba,Beaujean:2013soa,Crivellin:2015mga,Lyon:2014hpa} and references therein). It is still not clear if this discrepancy could be caused by an underestimation of the theory uncertainty on hadronic effects or if it requires a New Physics explanation.
Since short-distance effects should be universal in all $b\to s \mu\mu$ transitions, measuring other such transitions can shed light on this situation. Recently, the S-wave contribution to \BdToKpimm decays has been measured in the $644<\mkpi<1200\mevcc$ region~\cite{LHCb-PAPER-2016-012}. 

\begin{table}[!b]
%\caption{Expected resonant contributions above the \KstP mass range. For each, the spin-parity, $J^P$, and branching fraction to $\kaon\pion$, $\mathcal{B}(K\pi)$, are given (taken from Ref.~\cite{Lu:2011jm}).}
\caption{Expected resonant contributions above the \KstP mass range. For each, the spin-parity, $J^P$, and branching fraction to $\kaon\pion$, $\mathcal{B}(K\pi)$, are given (taken from Ref.~\cite{Olive:2016xmw}).}
\label{tab:introduction:states}
\centering
\begin{tabular}{c|c|c|c|r}
    Resonance & $J^{P}$ & Mass [$\mathrm{Me\kern -0.1em V\!/}c^2$] & Full width [$\mathrm{Me\kern -0.1em V\!/}c^2$]  & $\mathcal{B}(K\pi)~[\%]$ \\
   \hline
   $K^\ast(1410)^0$ & $1^{-}$& $\hphantom{0.}1414 \pm 15\hphantom{.}$& $232 \pm 21\hphantom{0}$  & $6.6 \pm 1.3$ \\
   $K^\ast_0(1430)^0$ & $0^{+}$ & $\hphantom{0.}1425 \pm 50\hphantom{.}$ & $270 \pm 80\hphantom{0}$ & $\hphantom{.}93 \pm 10\hphantom{.}$ \\
   $K^\ast_2(1430)^0$ & $2^{+}$ & $1432.4\pm 1.3$ & $109 \pm 5\hphantom{00}$ & $49.9 \pm 1.2$ \\
   $K^\ast(1680)^0$ & $1^{-}$ & $\hphantom{0.}1717 \pm 27\hphantom{.}$ & $322 \pm 110$ & $38.7 \pm 2.5$ \\
   $K^\ast_3(1780)^0$ & $3^{-}$ & $\hphantom{0.}1776 \pm 7\hphantom{0.}$ & $159 \pm 21\hphantom{0}$ & $18.8 \pm 1.0$ \\
   $K^\ast_4(2045)^0$ & $4^{+}$ & $\hphantom{0.}2045 \pm 9\hphantom{0.}$ & $198 \pm 30\hphantom{0}$ & $9.9 \pm 1.2$ \\
 \end{tabular}
 \end{table}

Since the dominant structures in the \kpi invariant mass spectrum of \BdToKpimm above the P-wave $\Kstar(892)^{0}$ are resonances in the 1430\mevcc region, this is a natural region to study. The relevant \Kstarz states above the \KstP mass range are listed in Table~\ref{tab:introduction:states}. Throughout this paper, the symbol $\Kstarz$ denotes any neutral strange meson in an excited state that decays to a \Kp\pim final state. In the 1430\mevcc region, contributions are expected from the S-wave $K^\ast_0(1430)^0$, P-wave $K^\ast(1410)^0$ and D-wave $K^\ast_2(1430)^0$ states, as well as the broad P-wave $K^\ast(1680)^0$ state. The mass region of the higher $K^\ast_J$ resonances was studied in Ref.~\cite{Lu:2011jm} with model-dependent theoretical predictions based on QCD form-factors. However, since the form-factors for broad resonances remain poorly known, a more model-independent prescription was provided in Ref.~\cite{spd-paper}, which is used in this analysis.

\begin{figure}[!tb]
 \centering
 \includegraphics[width=0.6\linewidth]{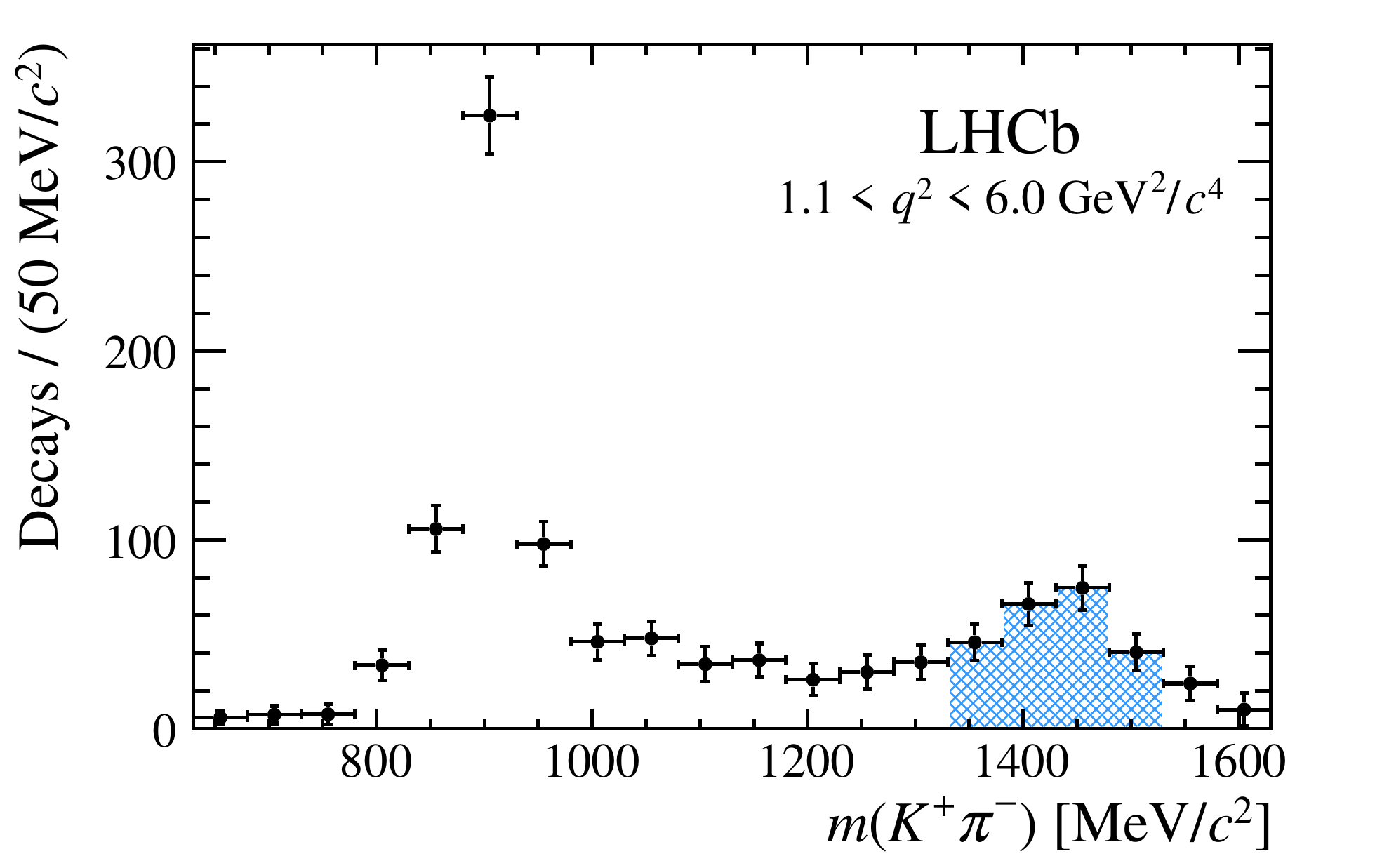}
 \caption{Background-subtracted \mkpi distribution for \BdToKpimm decays in the range $1.1<\qsq<6.0\gevgevcccc$. The region $1330<\mkpi<1530~\mevcc$ is indicated by the blue, hatched area.}
\label{fig:full-mkpi}
\end{figure}

The \mkpi distribution for \BdToKpimm decays in the range $1.1<\qsq<6.0\gevgevcccc$ and $630<\mkpi<1630\mevcc$ is shown in Fig.~\ref{fig:full-mkpi}, where $\qsq \equiv m^2(\mup \mun)$. The candidates are obtained using the selection described in Sec.~\ref{sec:selection} and the background component is subtracted using the \sPlot technique~\cite{Pivk:2004ty}. The main structures are observed around the mass of the $\Kstar(892)^{0}$ resonance and in the 1430\mevcc region. 

This paper presents the first measurements of the differential branching fraction and angular moments of \BdToKpimm in the region $1330<\mkpi<1530\mevcc$. The values of the differential branching fraction are reported in five bins of \qsq between 0.1 and 8.0\gevgevcccc, and in the range $1.1<\qsq<6.0\gevgevcccc$ for which the angular moments are also measured. The measurements are based on samples of $pp$ collisions collected by the \lhcb experiment in Run 1, corresponding to integrated luminosities of 1.0\invfb at a centre-of-mass energy of 7\tev and 2.0\invfb at 8\tev.

\section{Angular distribution}
\label{sec:angular-distribution}

\begin{figure}
\centering
\subfigure[]{
\centering
\includegraphics[width=0.47\textwidth]{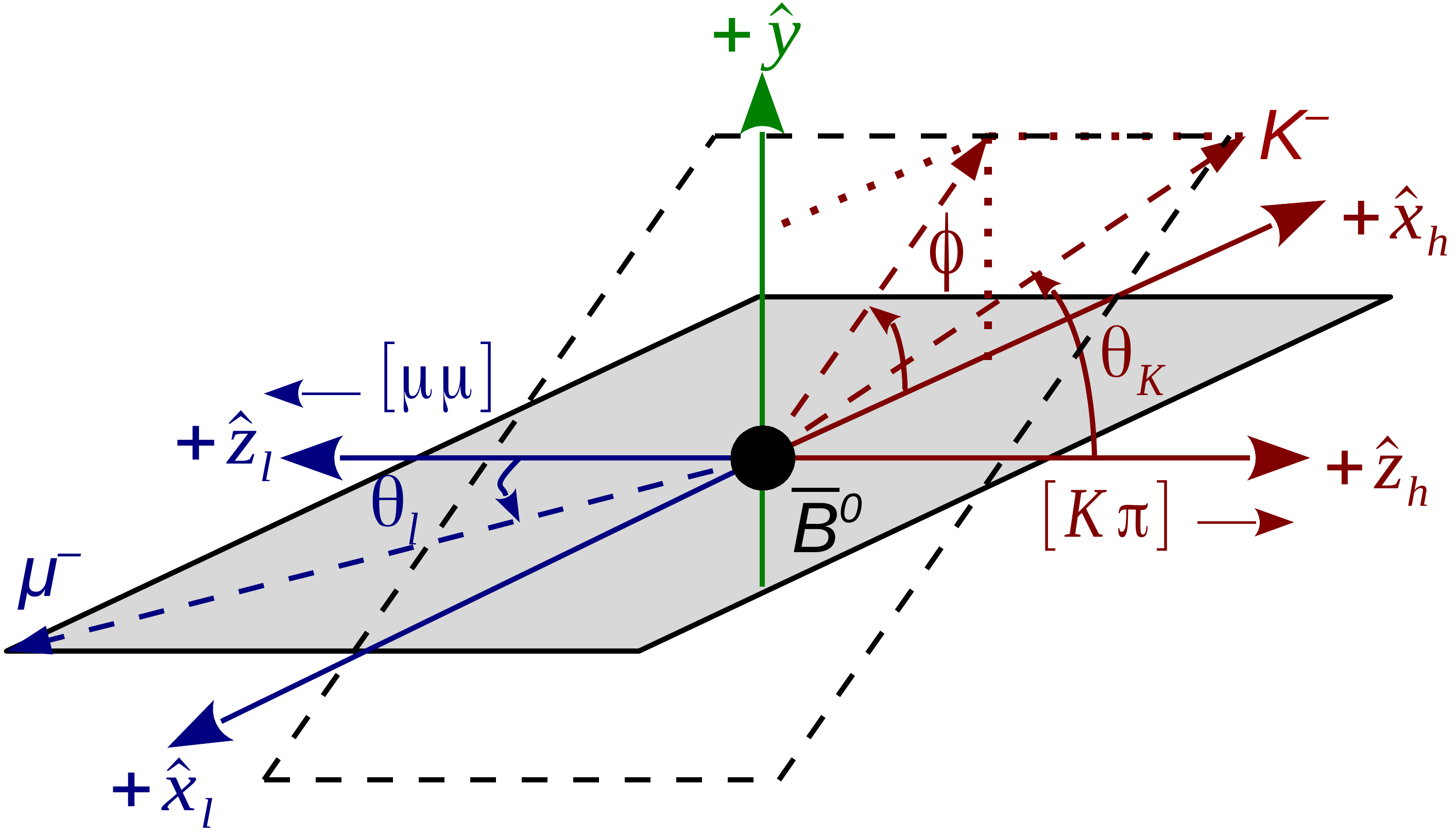}
}
\subfigure[]{
\centering
\includegraphics[width=0.47\textwidth]{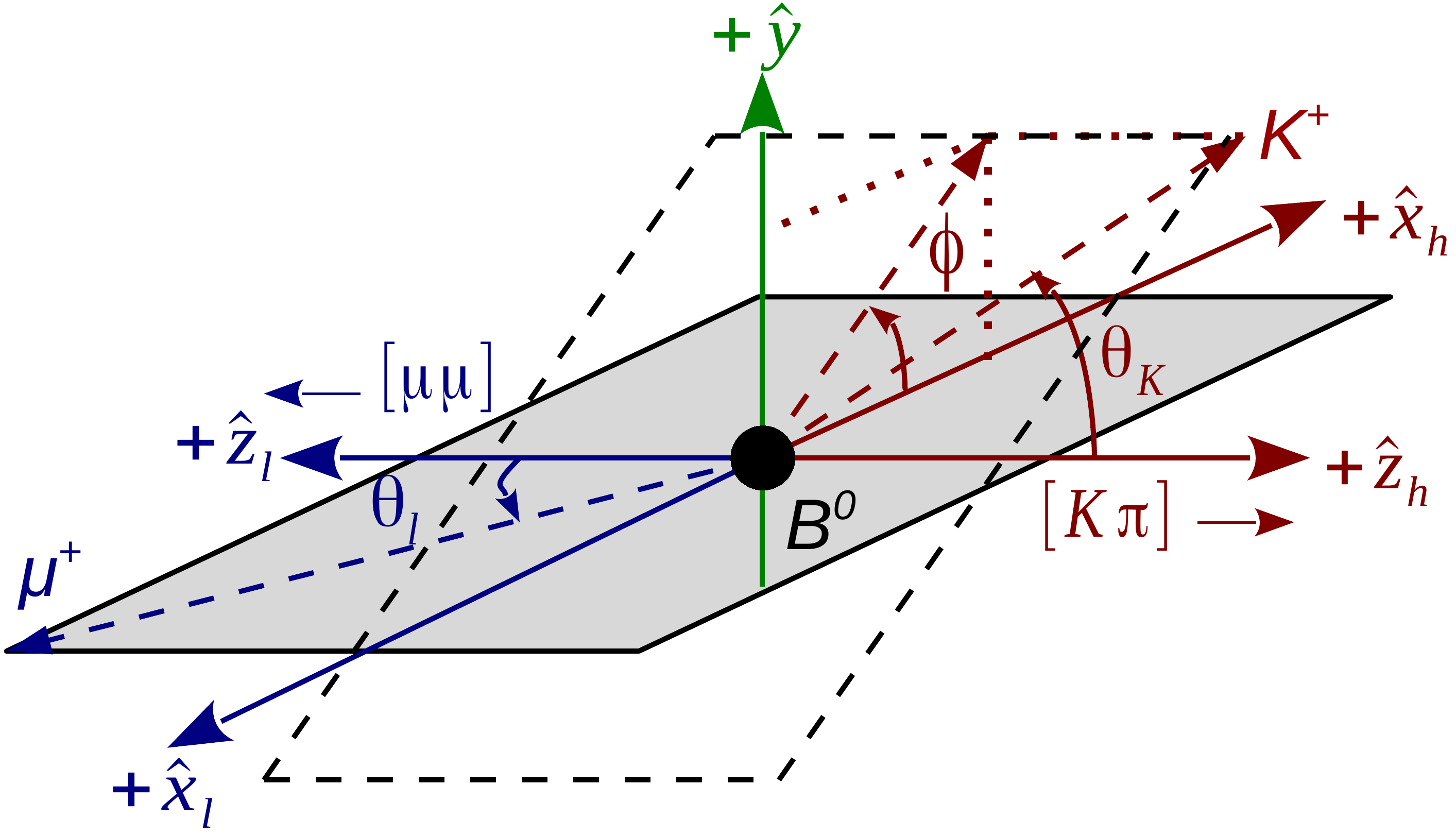}
}
\caption{Angle conventions for (a) $\Bzb \to \Km \pip \mun \mup$ and (b) $\Bz \to \Kp \pim \mup \mun$, as described in Ref.~\cite{spd-paper}. The leptonic and hadronic frames are back-to-back with a common $\hat{y}$ axis. For the dihedral angle $\phi$ between the leptonic and hadronic decay planes, there is an additional sign flip $\phi\to -\phi$ compared to previous \lhcb analyses~\cite{LHCb-PAPER-2011-020,LHCb-PAPER-2013-019,LHCb-PAPER-2013-037,LHCb-PAPER-2015-051}.
}
\label{fig:angle_conventions}
\end{figure}

The final state of the decay \BdToKpimm is fully described by five kinematic variables: three decay angles (\thetal, \thetak, $\phi$), \mkpi, and \qsq.
Figure~\ref{fig:angle_conventions}a shows the angle conventions for the $\Bzb$ decay (containing a \bquark quark): the back-to-back leptonic and hadronic systems share a common $\hat{y}$ axis and have opposite $\hat{x}$ and $\hat{z}$ axes. 
The negatively charged lepton is used to define the leptonic helicity angle $\thetal$ for the $\Bzb$.
The quadrant of the dihedral angle $\phi$ between the dimuon and the $\Kstarzb \to \Km\pip$ decay planes is determined by requiring the azimuthal angle of the $\mun$ to be zero in the leptonic helicity frame. The azimuthal angle of the $\Km$ in the hadronic helicity frame is then equal to $\phi$. Compared to the dihedral angle used in the previous \lhcb analyses~\cite{LHCb-PAPER-2011-020,LHCb-PAPER-2013-019,LHCb-PAPER-2013-037,LHCb-PAPER-2015-051}, there is a sign flip, $\phi \to -\phi$, in the convention used here. For the $\Bz$ decay (containing a \bquarkbar quark), the charge conjugation is performed explicitly, and the angles are shown in Fig.~\ref{fig:angle_conventions}b, where for the $\Bz$, the $\mup$ and $\Kp$ directions are used to define the angles. An additional minus sign is added to the dihedral angle when performing the \CP conjugation, in order to keep the measured angular observables the same between $\Bzb$ and $\Bz$ in the absence of direct \CP violation.

In the limit where \qsq is large compared to the square of the muon mass, the \CP-averaged differential decay rate of \BdToKpimm with the \Kp\pim system in a S-, P-, or D-wave configuration can be expanded in an orthonormal basis of angular functions $f_i(\Omega)$ as

\begin{equation}
\label{eqn:vector_moments}
\frac{\deriv\Gamma }{\deriv\qsq\,\deriv\Omega} \propto \sum^{41}_{i=1} f_i (\Omega) \Gamma_i(\qsq)
 \quad\mbox{with}\quad
\Gamma_i(\qsq) = \Gamma^L_i(\qsq) + \eta^{L\to R}_i\; \Gamma^R_i(\qsq),
\end{equation}

\noindent where $\deriv\Omega = \deriv\!\ctl\,\deriv\!\ctk\,\deriv\phi$, and $L$ and $R$ denote the (left- and right-handed) chirality of the lepton system~\cite{spd-paper}. The sign $\eta^{L\to R}_i=\pm 1$ depends on whether $f_i$ changes sign under $\thetal \to \pi + \thetal$. 
\noindent The orthonormal angular basis is constructed out of spherical harmonics, \mbox{$Y^m_l \equiv Y^m_l (\thetal,\phi)$}, and reduced spherical harmonics, \mbox{${P^m_l \equiv \sqrt{2 \pi}Y^m_l(\thetak,0)}$}.
 
The transversity-basis moments of the 41 orthonormal angular functions are given in Appendix~\ref{sec:appendix:angular-distribution}. 
The convention is that the amplitudes correspond to the $\Bzb$ decay, with the corresponding amplitudes for the $\Bz$ decay obtained by flipping the signs of the helicities and weak phases. The S-, P- and D-wave transversity amplitudes are denoted as $S^{\{L,R\}}$, $H^{\{L,R\}}_{\{0,\parallel,\perp\}}$ and $D^{\{L,R\}}_{\{0,\parallel,\perp\}}$, respectively. 

The measured angular observables are averaged over the range $1330<\mkpi<1530~\mevcc$ and $1.1<\qsq<6.0\gevgevcccc$. This \qsq range is part of the large-recoil regime where the recoiling \Kstarz has a relatively large energy, $E_{\Kstarz}$, as measured in the rest frame of the parent \B meson. In the limit $\Lambda_{\rm QCD}/E_{\Kstarz} \to 0$, the uncertainties arising from hadronic effects in the relevant form-factors are reduced at leading order, resulting in more reliable theory predictions~\cite{DescotesGenon:2013wba}. The high-$\qsq$ region above the $\psi(2S)$ resonance is polluted by broad charmonium resonances and is also phase-space suppressed for higher \mkpi masses. Therefore, that region is not considered in this study.

In the present analysis, the first moment, $\Gamma_{1}(\qsq)$, corresponds to the total decay rate. From this, 40 normalised moments for $i \in \{2,...,41\}$ are defined as
\begin{equation}
\label{eqn:norm_mom_def}
\overline{\Gamma}_i(\qsq) = \frac{\Gamma_{i}(\qsq)}{\Gamma_{1}(\qsq)}.
\end{equation}
\noindent These form the set of observables that are measured in the angular moments analysis described in Sec.~\ref{sec:angular-analysis}.

\section{Detector and simulation}
\label{sec:Detector}

The \lhcb detector~\cite{Alves:2008zz,LHCb-DP-2014-002} is a single-arm forward
spectrometer covering the \mbox{pseudorapidity} range $2<\eta <5$,
designed for the study of particles containing \bquark or \cquark
quarks. The detector includes a high-precision tracking system
consisting of a silicon-strip vertex detector surrounding the $pp$
interaction region, a large-area silicon-strip detector located
upstream of a dipole magnet with a bending power of about
$4{\mathrm{\,Tm}}$, and three stations of silicon-strip detectors and straw
drift tubes placed downstream of the magnet.
The tracking system provides a measurement of momentum of charged particles with
a relative uncertainty that varies from 0.5\% at low momentum to 1.0\% at 200\gevc.
The minimum distance of a track to a primary vertex (PV), the impact parameter (IP), 
is measured with a resolution of $(15+29/\pt)\mum$,
where \pt is the component of the momentum transverse to the beam, in\,\gevc.
Different types of charged hadrons are distinguished using information
from two ring-imaging Cherenkov detectors. 
Photons, electrons and hadrons are identified by a calorimeter system consisting of
scintillating-pad and preshower detectors, an electromagnetic
calorimeter and a hadronic calorimeter. Muons are identified by a
system composed of alternating layers of iron and multiwire
proportional chambers.
The online event selection is performed by a trigger, 
which consists of a hardware stage, based on information from the calorimeter and muon
systems, followed by a software stage, which applies a full event
reconstruction.

Simulated signal events are used to determine the effect of the detector geometry, trigger, reconstruction and selection on the angular distributions of the signal and of the \BdToJPsiKstP mode, which is used for normalisation. Additional simulated samples are used to estimate the contribution from specific background processes.
In the simulation, $pp$ collisions are generated using
\pythia~\cite{Sjostrand:2006za,Sjostrand:2007gs} 
 with a specific \lhcb
configuration~\cite{LHCb-PROC-2010-056}.  Decays of hadronic particles
are described by \evtgen~\cite{Lange:2001uf}, in which final-state
radiation is generated using \photos~\cite{Golonka:2005pn}. The
interaction of the generated particles with the detector, and its response,
are implemented using the \geant
toolkit~\cite{Allison:2006ve} as described in
Ref.~\cite{LHCb-PROC-2011-006}.
Data-driven techniques are used to correct the simulation for mismodelling of the detector occupancy, the \Bz meson momentum and vertex quality distributions, and particle identification performance.

\section{Selection of signal candidates}
\label{sec:selection}

The \BdToKpimm signal candidates are first required to pass the hardware trigger,
which selects events containing at least one muon with transverse momentum $\pt>1.48\gevc$
in the 7\tev data or $\pt>1.76\gevc$ in the 8\tev data.
In the subsequent software trigger, at least
one of the final-state particles is required to have both
$\pt>1.0\gevc$ and an impact parameter larger than $100\mum$ with respect to all PVs in the
event. Finally, the tracks of two or more of the final-state
particles are required to form a vertex significantly
displaced from all PVs.

Signal candidates are formed from a pair of oppositely charged tracks identified as muons, combined with two oppositely charged tracks identified as a kaon and a pion. These signal candidates are then required to pass a set of loose preselection requirements, identical to those described in Ref.~\cite{LHCb-PAPER-2015-051} with the exception that the $\Kp\pim$ system is permitted to be in the wider mass range $630<\mkpi<1630\mevcc$.  This allows the decay \BdToJPsiKstP to be used as a normalisation mode for the branching fraction measurement. 
Candidates are required to have good quality vertex and track fits, and a reconstructed $\Bz$ invariant mass in the range $5170<\mkpimm<5700\mevcc$.
From this point onwards, the normalisation mode is selected in the range $796<\mkpi<996\mevcc$ and the signal in the range $1330<\mkpi<1530\mevcc$.
 
The backgrounds from combining unrelated particles, mainly from different \bquark and \cquark hadron decays, are referred to as combinatorial. Such backgrounds are suppressed with the use of a Boosted Decision Tree~(BDT)~\cite{Breiman,AdaBoost}. 
The BDT used for the present analysis is identical to that described in Ref.~\cite{LHCb-PAPER-2015-051} and the same working point is used.

Exclusive background processes can mimic the signal if their final states are misidentified or misreconstructed.  
For the present analysis, the requirements of Ref.~\cite{LHCb-PAPER-2015-051} for the $\Kstar(892)^{0}$ region are applied to a wider \mkpi invariant mass window. However, to reduce the expected contamination from peaking background to the level of 2\% of the signal yield, it is necessary to modify two of them. First, the requirement to remove contributions from \BdToJPsiKstP candidates, where the \pim (\Kp) is misidentified as a \mun (\mup) and the \mun (\mup) is misidentified as a \pim (\Kp), is tightened by extending the invariant mass window of the $\mup\pim$ ($\Kp\mun$) system and requiring stricter muon identification criteria.  Second, the requirement to remove the contributions from genuine \BdToKpimm decays where the two hadron hypotheses are interchanged is tightened by requiring stricter hadron identification criteria.

\section{Acceptance correction}
\label{sec:acceptance}

The triggering, reconstruction and selection of candidates distorts their kinematic distributions.
The dominant acceptance effects are due to the requirements on track momentum and impact parameter.
 
\begin{figure}[!tb]
 \centering
 \includegraphics[width=0.49\linewidth]{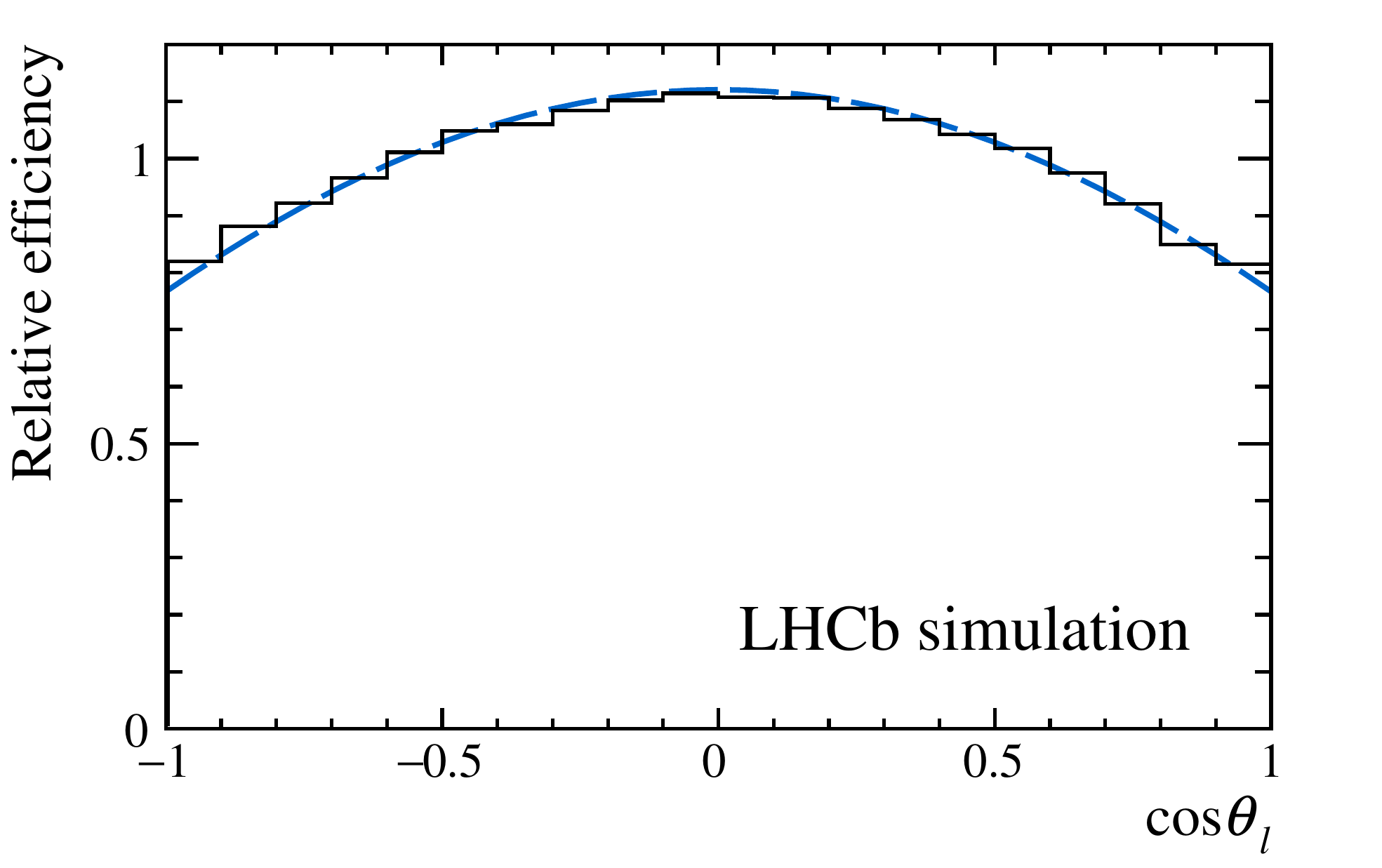}
 \includegraphics[width=0.49\linewidth]{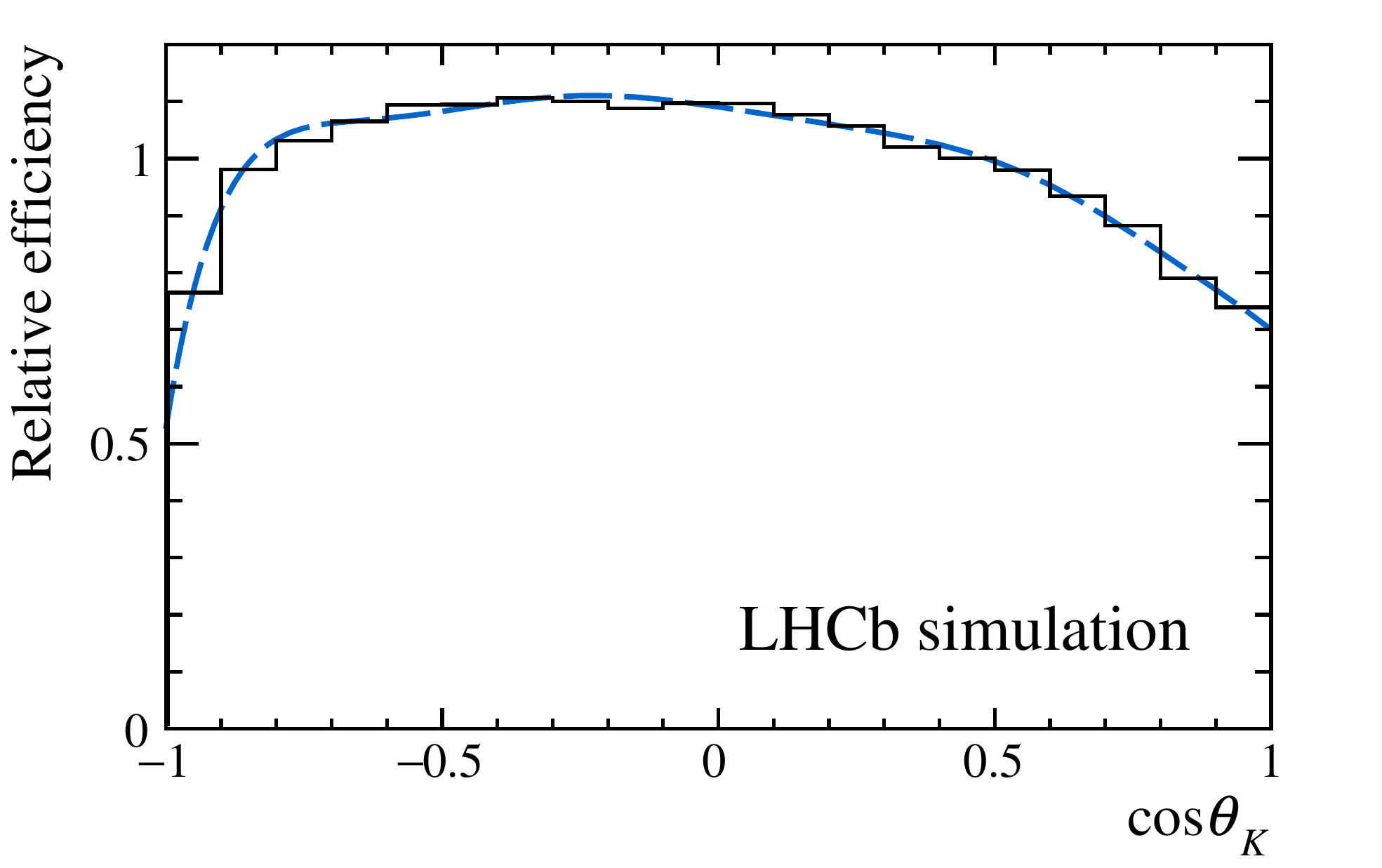}\\
 \includegraphics[width=0.49\linewidth]{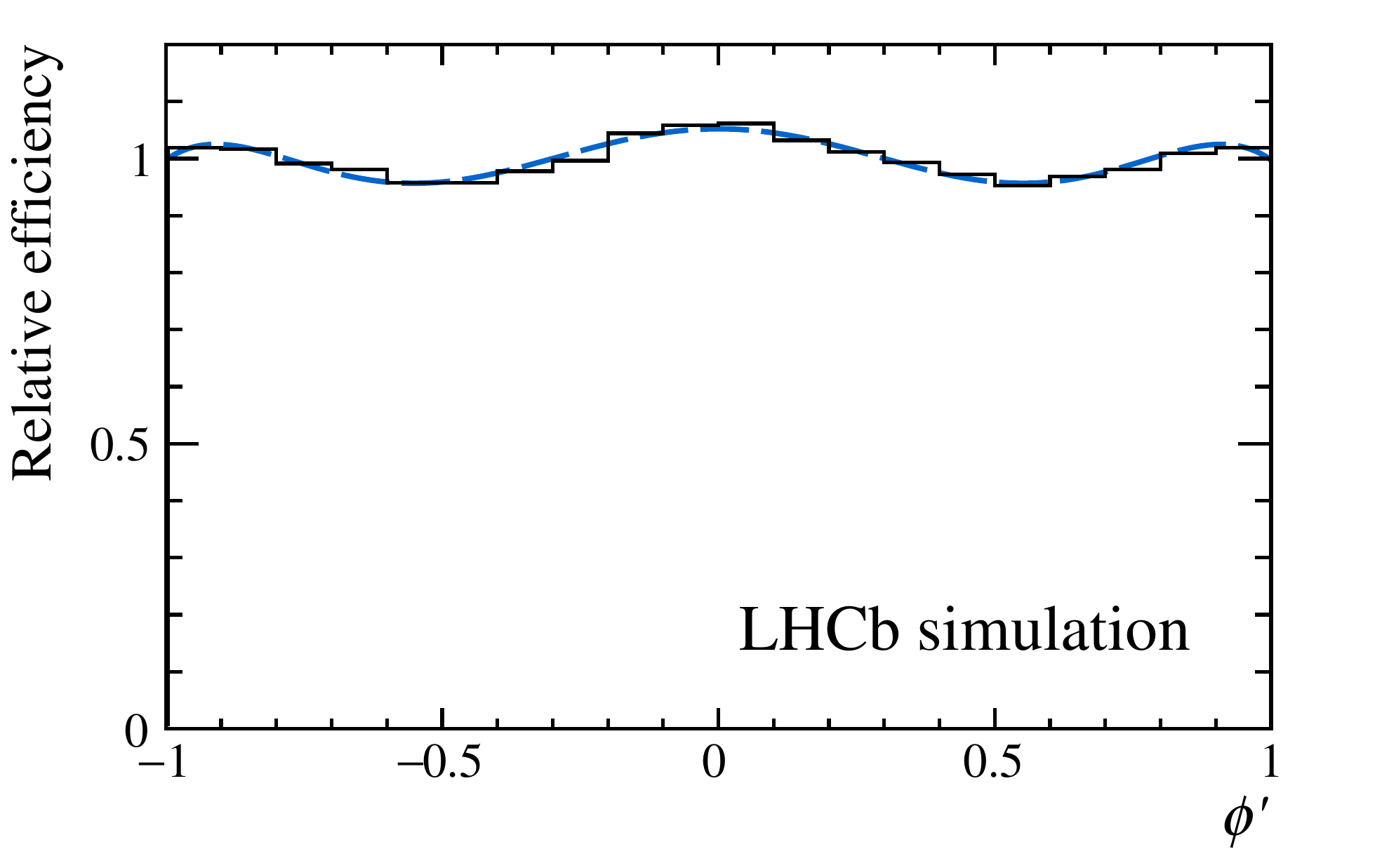}
 \caption{Relative efficiency in \ctl, \ctk and \phip in the region $1.1<\qsq<6.0\gevgevcccc$ and $1330<\mkpi<1530\mevcc$ as determined from a moment analysis of simulated \BdToKpimm decays, shown as a histogram. The efficiency function is shown by the blue, dashed line.}
\label{fig:acceptance}
\end{figure}

The method for obtaining the acceptance correction, described in Ref.~\cite{LHCb-PAPER-2015-051}, is extended to include the \mkpi dimension.
The efficiency is parameterised in terms of Legendre polynomials of order $n$, $L_n(x)$, as
\begin{equation}
\begin{split}
 \varepsilon(\qsqp\!,\ctl,\ctk,\phip&,\mkpip) = \\
 & \sum_{hijkl} c_{hijkl}\,L_{h}(\qsqp)\,L_{i}(\ctl)\,L_{j}(\ctk)\,L_{k}(\phip)\,L_{l}(\mkpip).
  \end{split}
  \label{eqn:legendre}
\end{equation}
As the polynomials are defined over the domain $x\in[-1,1]$, the variables \qsqp, \phip and \mkpip are used, which are obtained by linearly transforming \qsq, $\phi$ and \mkpi to lie in this range.
The sum in Eq.~\ref{eqn:legendre} encompasses $L_n(x)$ up to fourth order in \ctl and \mkpip, sixth order in \phip and \qsqp, and eighth order in \ctk.  The coefficients $c_{hijkl}$ are determined using a moment analysis of simulated \BdToKpimm decays, generated according to a phase space distribution. The angular acceptance as a function of \ctl, \ctk and \phip in the region $1.1<\qsq<6.0\gevgevcccc$ and $1330<\mkpi<1530\mevcc$ is shown in Fig.~\ref{fig:acceptance}.

\section{The \boldmath{\mkpimm} invariant mass distribution}
\label{sec:massfit}

\begin{figure}[!tb]
 \centering
 \includegraphics[width=0.48\linewidth]{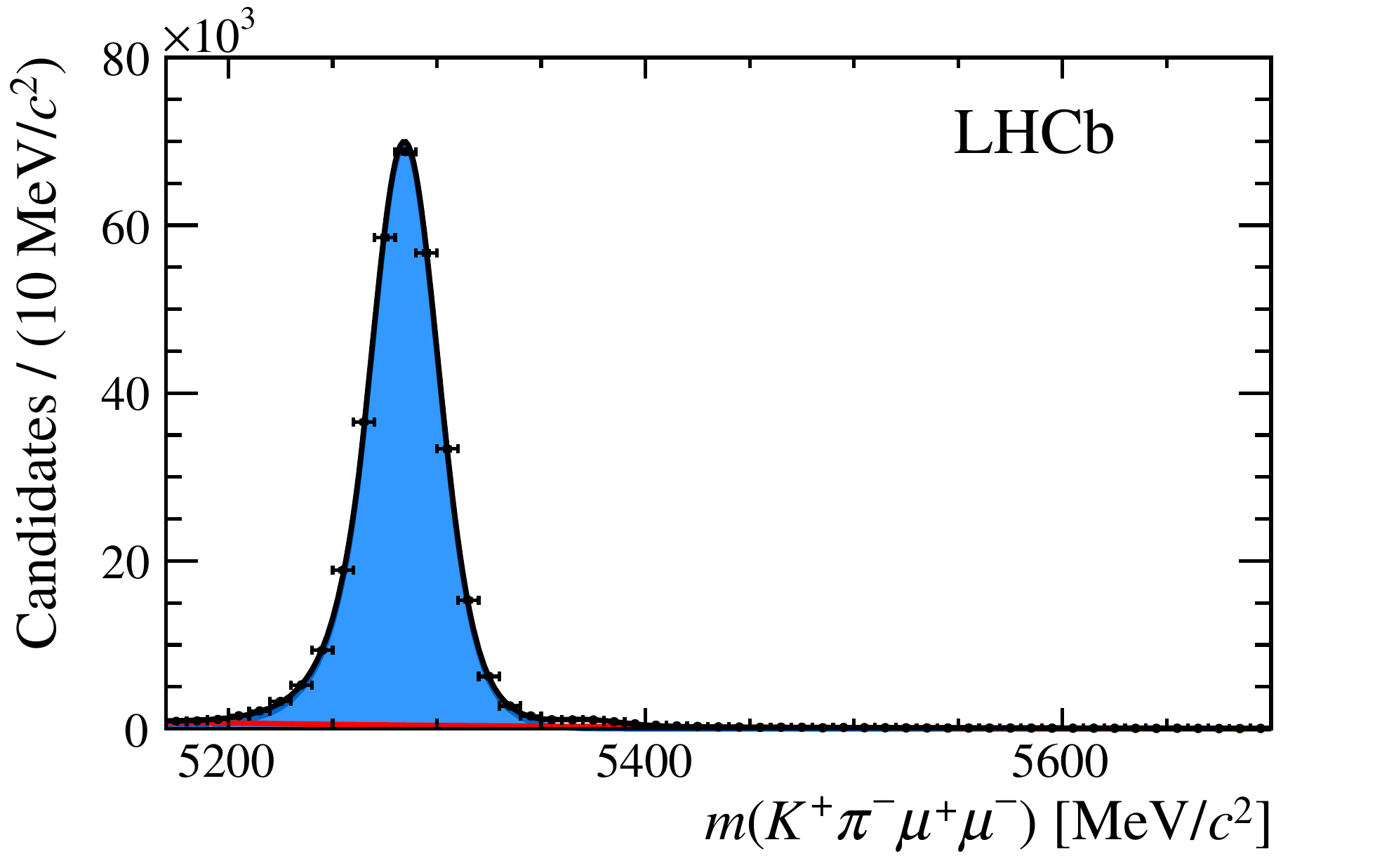}
 \includegraphics[width=0.48\linewidth]{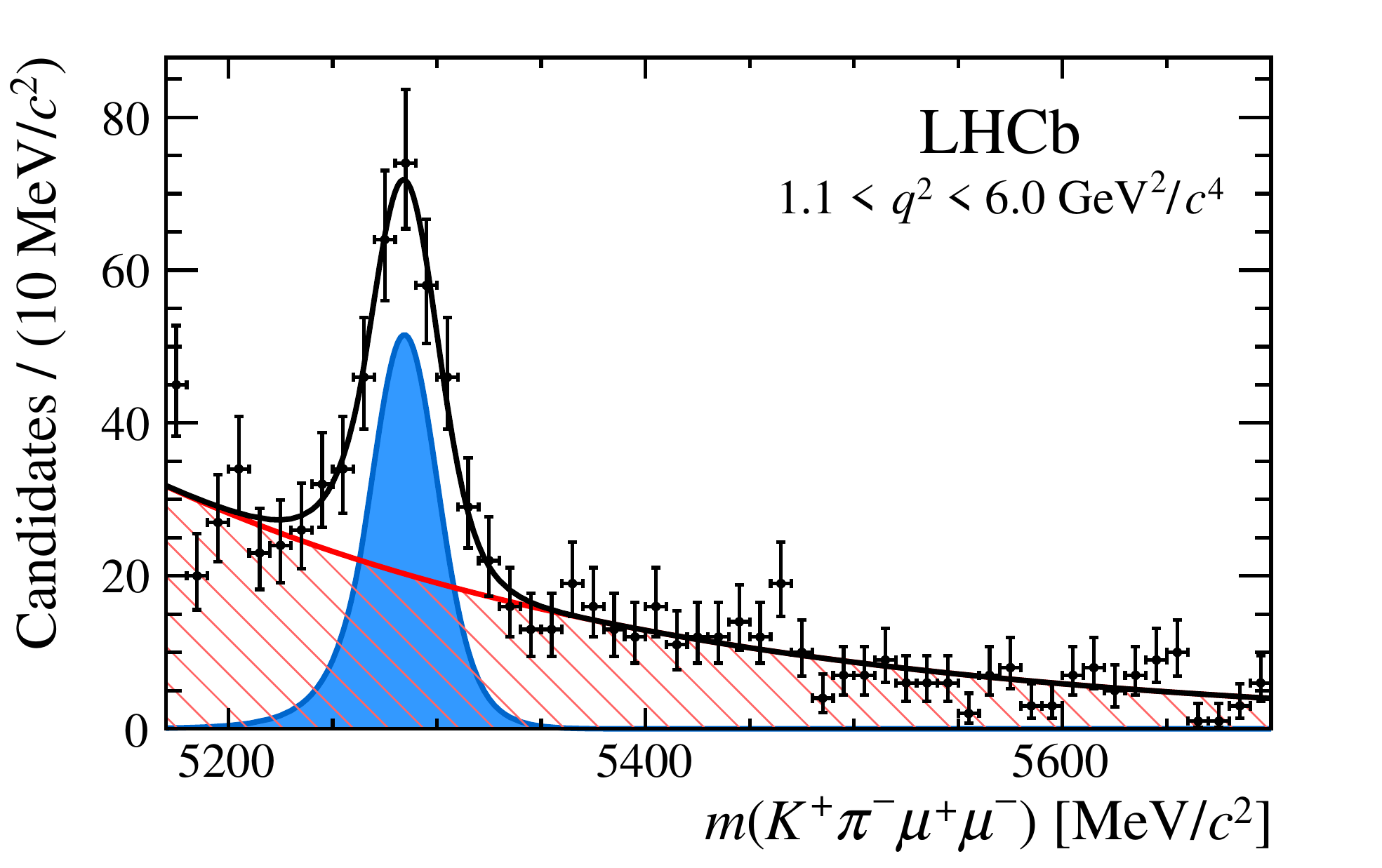}
 \caption{Invariant mass \mkpimm for (left) the control decay \BdToJPsiKst and (right) the signal decay \BdToKpimm in the bin $1.1 < \qsq < 6.0\gevgevcccc$. The solid black line represents the total fitted function.  The individual components of the signal (blue, shaded area) and combinatorial background (red, hatched area) are also shown.}
\label{fig:massfit}
\end{figure}

The invariant mass $m(\Kp\pim\mumu)$ is used to discriminate between signal and background. The signal distribution is modelled as the sum of two Gaussian functions with a common mean, each with a power-law tail on the low-mass side. The parameters describing the shape of the mass distribution of the signal are determined from a fit to the \BdToJPsiKstP control mode, as shown in Fig.~\ref{fig:massfit}, and are subsequently fixed when fitting the \BdToKpimm candidates. An additional component is included in the fit to the control mode to model the contribution from \BsToJPsiKst decays. A single scaling factor is used to correct the width of the Gaussian functions to account for variations in the shape of the mass distribution of the signal observed in simulation, due to the different regions of \mkpi and \qsq between the control mode and signal mode. The combinatorial background is modelled using an exponential function.
The fit to \BdToKpimm candidates in the range $1.1 < \qsq < 6.0\gevgevcccc$ is shown in Fig.~\ref{fig:massfit}. The signal yield in the range $1.1 < \qsq < 6.0\gevgevcccc$ is $229 \pm 21$. The fits to \BdToKpimm candidates in each of the \qsq bins used for the differential branching fraction measurement are shown in Appendix~\ref{sec:appendix:massfit}.

\section{Differential branching fraction}
\label{sec:differential-bf}

The differential branching fraction $\deriv\BF/\deriv\qsq$ of the decay \BdToKpimm in an interval ($q^{2}_{\text{min}}$, $q^{2}_{\text{max}}$) is given by

\begin{equation}
\begin{split}
\frac{\deriv\BF}{\deriv\qsq} = \frac{1}{(q^{2}_{\text{max}} - q^{2}_{\text{min}}) }&f_{\KstP}\BF(\BdToJPsiKstP)\BF(\decay{\jpsi}{\mumu}) \\
\times&\BF(\decay{\KstP}{\Kp\pim})\frac{N'_{\kpimm}}{(1-F_{\rm S}^{\jpsi\Kstarz})N'_{\jpsi\Kstarz}},
\end{split}
\label{eqn:dbfdq2}
\end{equation}

\noindent where $N'_{\kpimm}$ and $N'_{\jpsi\Kstarz}$ are the acceptance-corrected yields of the \BdToKpimm and ${\Bz \to \jpsi(\to \mup \mun)\Kstarz(\to \Kp \pim)}$ decays, respectively. The \BdToJPsiKst yield has to be corrected for the S-wave fraction within the $796<\mkpi<996~\mevcc$ window of \BdToJPsiKst decays, $F_{\rm S}^{\jpsi\Kstarz}$. The value of $F_{\rm S}^{\jpsi\Kstarz}=0.084\pm 0.01$ is obtained from Ref.~\cite{LHCb-PAPER-2013-023}, after recalculation for the $\mkpi$ range $796<\mkpi<996~\mevcc$. The branching fractions $\BF(\BdToJPsiKstP)$, $\BF(\decay{\jpsi}{\mumu})$ and $\BF(\decay{\KstP}{\Kp\pim})$ are $(1.19\pm0.01\pm0.08)\times10^{-3}$~\cite{Chilikin:2014bkk}, $(5.961 \pm 0.033) \times 10^{-2}$~\cite{Olive:2016xmw} and 2/3, respectively. The fraction $f_{\KstP}$ is used to scale the value of $\BF(\BdToJPsiKstP)$ to the appropriate \mkpi range and is calculated by integrating the $\KstP$ line shape given in Ref.~\cite{Chilikin:2014bkk} over the range $796<\mkpi<996~\mevcc$.

In order to obtain the acceptance-corrected yield, the efficiency function described in Sec.~\ref{sec:acceptance} is used to evaluate an acceptance weight for each candidate. An average acceptance weight is determined for both the \BdToJPsiKst candidates and the signal candidates in each \qsq bin. The acceptance-corrected yield is then equal to the measured yield multiplied by the average weight. The average weight is calculated within the $\pm 50\mevcc$ signal window around the mean \Bz mass and also in the background region taken from the upper mass sideband in the range $5350<\mkpimm<5700\mevcc$. The latter is subsequently used to subtract the background contribution from the average weight obtained in the $\pm 50\mevcc$ window, taking into account the extrapolated background yield in this window. This method avoids making any assumption about the unknown angular distribution of the \BdToKpimm decay.

\begin{figure}[!tb]
 \centering
 \includegraphics[width=0.6\textwidth]{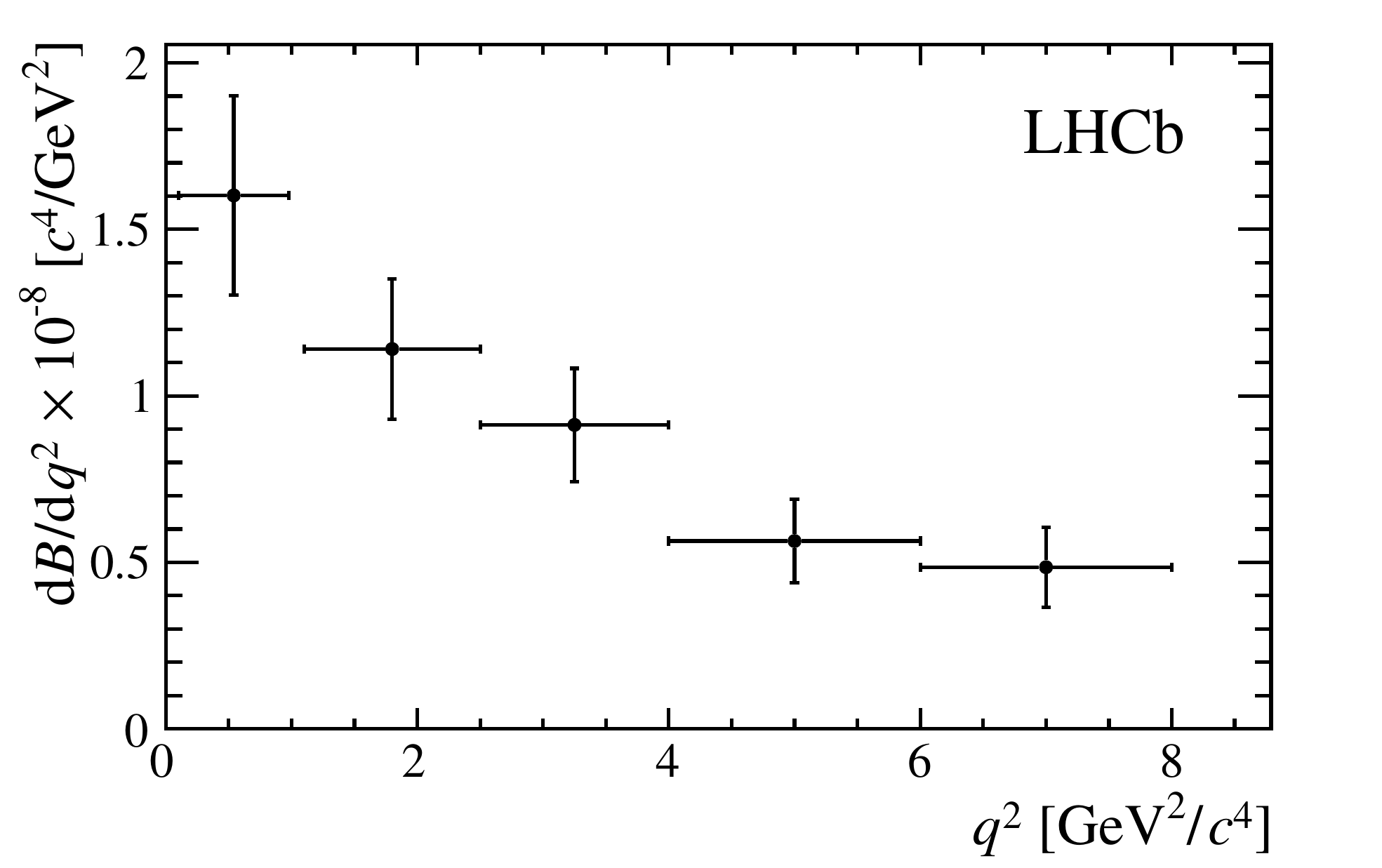}
 \caption{Differential branching fraction of \BdToKpimm in bins of \qsq for the range $1330<\mkpi<1530\mevcc$. The error bars indicate the sums in quadrature of the statistical and systematic uncertainties.}
 \label{fig:bf}
\end{figure}

The results for the differential branching fraction are given in Fig.~\ref{fig:bf}.  The uncertainties shown are the sums in quadrature of the statistical and systematic uncertainties.  The results are also presented in Table~\ref{tab:bf}.  The various sources of the systematic uncertainties are described in Sec.~\ref{sec:systematics}.

\begin{table}[!tb]
\caption{Differential branching fraction of \BdToKpimm in bins of \qsq for the range $1330<\mkpi<1530\mevcc$. The first uncertainty is statistical, the second systematic and the third due to the uncertainty on the \BdToJPsiKstP and $\decay{\jpsi}{\mumu}$ branching fractions.}
\label{tab:bf}
\begin{center}
\begin{tabular}{lc}
\qsq [$\gevgevcccc$] & $\deriv\BF/\deriv\qsq \times 10^{-8}~[c^{4}/\gev^{2}]$ \\
\hline
$[0.10,0.98]$ & 1.60 $\pm$ 0.28 $\pm$ 0.04 $\pm$ 0.11 \\
$[1.10,2.50]$ & 1.14 $\pm$ 0.19 $\pm$ 0.03 $\pm$ 0.08 \\
$[2.50,4.00]$ & 0.91 $\pm$ 0.16 $\pm$ 0.03 $\pm$ 0.06 \\
$[4.00,6.00]$ & 0.56 $\pm$ 0.12 $\pm$ 0.02 $\pm$ 0.04 \\
$[6.00,8.00]$ & 0.49 $\pm$ 0.11 $\pm$ 0.01 $\pm$ 0.03 \\
\hline
$[1.10,6.00]$ & 0.82 $\pm$ 0.09 $\pm$ 0.02 $\pm$ 0.06 \\
\end{tabular}
\end{center}
\end{table}

\section{Angular moments analysis}
\label{sec:angular-analysis}

The angular observables defined in Sec.~\ref{sec:angular-distribution} are determined using a moments analysis of the angular distribution, as outlined in Ref.~\cite{spd-paper}. This approach has the advantage of producing stable measurements with well-defined uncertainties even for small data samples. Similar methods using angular moments are described in Refs.~\cite{Beaujean:2015xea,Gratrex:2015hna}.

The 41 background-subtracted and acceptance-corrected moments are estimated as
\begin{equation}
 \label{eqn:moments}
\Gamma_i =  \sum_{k=1}^{n_{\rm sig}} w_{k}f_i(\Omega_k)  - x\sum_{k=1}^{n_{\rm bkg}} w_{k}f_i(\Omega_k)\\
\end{equation}

\noindent and the corresponding covariance matrix is estimated as

\begin{equation}
 \label{eqn:covariance}
 C_{ij} = \sum_{k=1}^{n_{\rm sig}} w^{2}_{k}f_i(\Omega_k)f_j(\Omega_k)   + x^2\sum_{k=1}^{n_{\rm bkg}} w^{2}_{k}f_i(\Omega_k)f_j(\Omega_k).
\end{equation}

\noindent Here $n_{\rm sig}$ and $n_{\rm bkg}$ correspond to the candidates in the signal and background regions, respectively. The signal region is defined within $\pm 50\mevcc$ of the mean \Bz mass, and the background region in the range $5350<\mkpimm<5700$~\mevcc.  The scale factor $x$ is the ratio of the estimated number of background candidates in the signal region over the number of candidates in the background region and is used to normalise the background subtraction.
It has been checked in data that the angular distribution of the background is independent of \mkpimm within the precision of this measurement, and that the uncertainty on $x$ has negligible impact on the results.
The weights, $w_{k}$, are the reciprocals of the candidates' efficiencies and account for the acceptance, described in Sec.~\ref{sec:acceptance}.

The covariance matrix describing the statistical uncertainties on the 40 normalised moments is computed as
\begin{align}
\label{eqn:red_cov_def}
\overline{C}_{ij} = \left[C_{ij} + \frac{\Gamma_i \Gamma_j}{\Gamma_1^2} C_{11} - \frac{\Gamma_i C_{1j} + \Gamma_j C_{1i}}{\Gamma_1}\right] \frac{1}{\Gamma_1^2},  \;\; i,j \in \{2,...,41\}.
\end{align}

\begin{figure}[!tb]
\centering
  \includegraphics[width=0.7\textwidth]{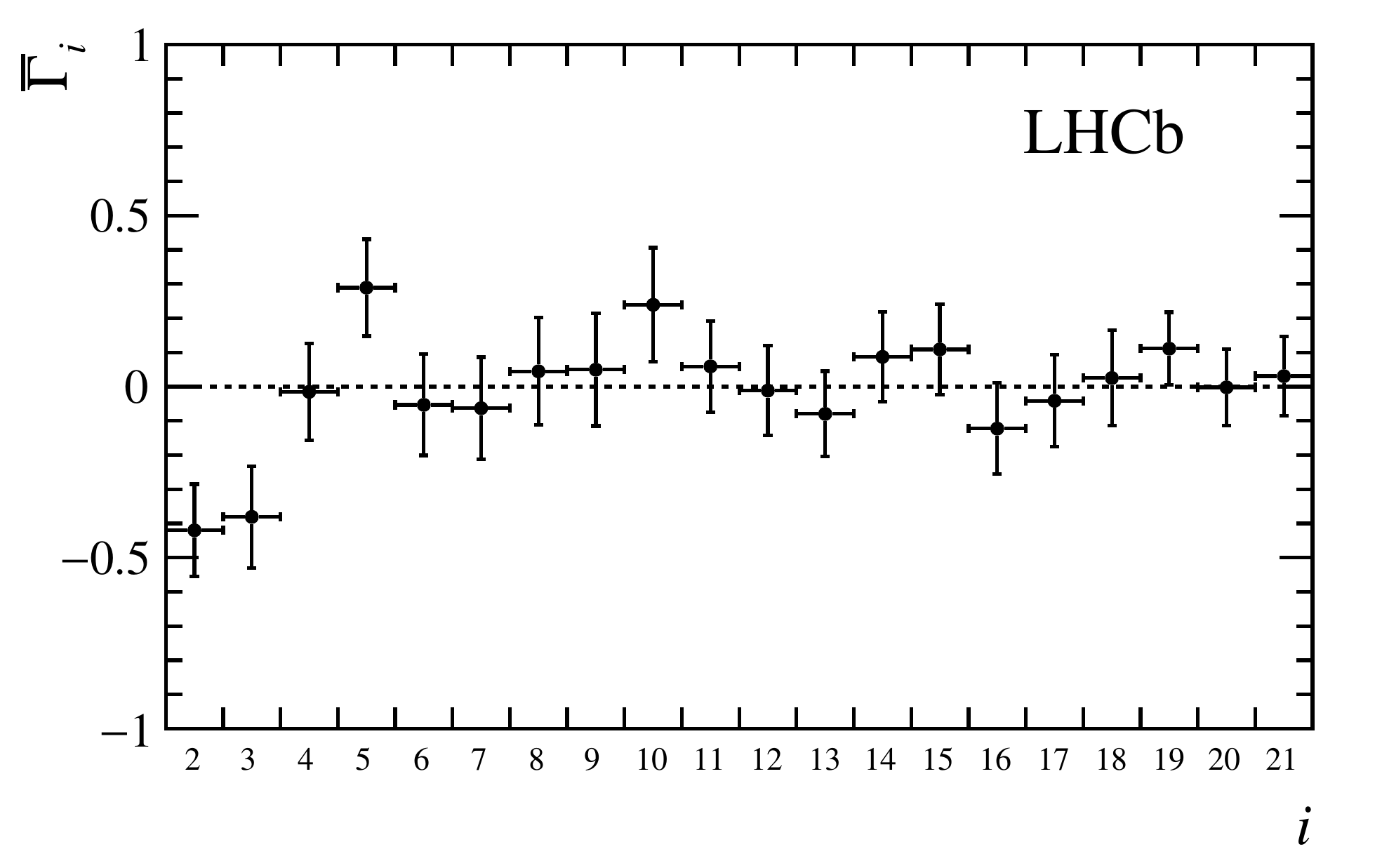}
  \includegraphics[width=0.7\textwidth]{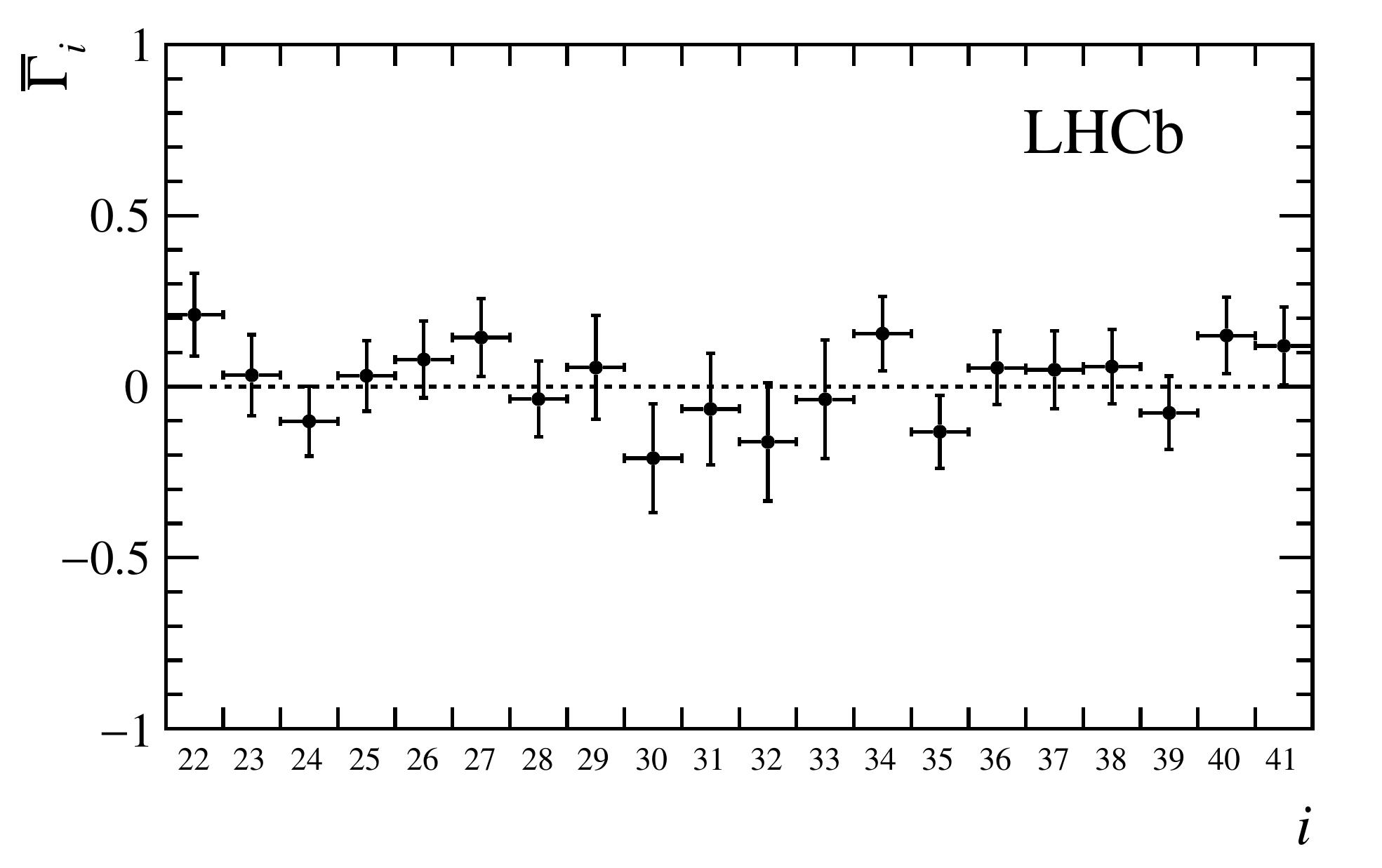}
  \caption{Measurement of the normalised moments, $\overline{\Gamma}_{i}$, of the decay \BdToKpimm in the range $1.1<\qsq<6.0\gevgevcccc$ and $1330<\mkpi<1530\mevcc$. The error bars indicate the sums in quadrature of the statistical and systematic uncertainties.}
  \label{fig:results:moments}
\end{figure}

The results for the normalised moments, $\overline{\Gamma}_{i}$, are given in Fig.~\ref{fig:results:moments}. The uncertainties shown are the sums in quadrature of the statistical and systematic uncertainties. The results are also presented in Table~\ref{tab:results:moments}. The various sources of the systematic uncertainties are described in Sec.~\ref{sec:systematics}. The complete set of numerical values for the measured moments and the covariance matrix is provided in Ref.~\cite{hepdata}.

\begin{table}[!tb]
\caption{Measurement of the normalised moments, $\overline{\Gamma}_{i}$, of the decay \BdToKpimm in the range $1.1<\qsq<6.0\gevgevcccc$ and $1330<\mkpi<1530\mevcc$. The first uncertainty is statistical and the second systematic.}
\label{tab:results:moments}
\centering
\begin{tabular}{l|c}
$\overline{\Gamma}_{i}$ & Value \\ 
\hline
$\overline{\Gamma}_{2}$ & $-0.42$ $\pm$ 0.13 $\pm$ 0.03 \\ 
$\overline{\Gamma}_{3}$ & $-0.38$ $\pm$ 0.15 $\pm$ 0.01 \\ 
$\overline{\Gamma}_{4}$ & $-0.02$ $\pm$ 0.14 $\pm$ 0.01 \\ 
$\overline{\Gamma}_{5}$ & \hphantom{$-$}0.29 $\pm$ 0.14 $\pm$ 0.02 \\ 
$\overline{\Gamma}_{6}$ & $-0.05$ $\pm$ 0.14 $\pm$ 0.04 \\ 
$\overline{\Gamma}_{7}$ & $-0.06$ $\pm$ 0.15 $\pm$ 0.03 \\ 
$\overline{\Gamma}_{8}$ & \hphantom{$-$}0.04 $\pm$ 0.16 $\pm$ 0.01 \\ 
$\overline{\Gamma}_{9}$ & \hphantom{$-$}0.05 $\pm$ 0.16 $\pm$ 0.02 \\ 
$\overline{\Gamma}_{10}$ & \hphantom{$-$}0.24 $\pm$ 0.17 $\pm$ 0.02 \\ 
$\overline{\Gamma}_{11}$ & \hphantom{$-$}0.06 $\pm$ 0.13 $\pm$ 0.01 \\ 
$\overline{\Gamma}_{12}$ & $-0.01$ $\pm$ 0.13 $\pm$ 0.02 \\ 
$\overline{\Gamma}_{13}$ & $-0.08$ $\pm$ 0.12 $\pm$ 0.01 \\ 
$\overline{\Gamma}_{14}$ & \hphantom{$-$}0.09 $\pm$ 0.13 $\pm$ 0.01 \\ 
$\overline{\Gamma}_{15}$ & \hphantom{$-$}0.11 $\pm$ 0.13 $\pm$ 0.00 \\ 
$\overline{\Gamma}_{16}$ & $-0.12$ $\pm$ 0.13 $\pm$ 0.01 \\ 
$\overline{\Gamma}_{17}$ & $-0.04$ $\pm$ 0.13 $\pm$ 0.01 \\ 
$\overline{\Gamma}_{18}$ & \hphantom{$-$}0.03 $\pm$ 0.14 $\pm$ 0.01 \\ 
$\overline{\Gamma}_{19}$ & \hphantom{$-$}0.11 $\pm$ 0.11 $\pm$ 0.01 \\ 
$\overline{\Gamma}_{20}$ & $-0.00$ $\pm$ 0.11 $\pm$ 0.01 \\ 
$\overline{\Gamma}_{21}$ & \hphantom{$-$}0.03 $\pm$ 0.12 $\pm$ 0.01 \\ 
\end{tabular}
\hspace{1em}
\begin{tabular}{l|c}
$\overline{\Gamma}_{i}$ & Value \\ 
\hline
$\overline{\Gamma}_{22}$ & \hphantom{$-$}0.21 $\pm$ 0.12 $\pm$ 0.01 \\ 
$\overline{\Gamma}_{23}$ & \hphantom{$-$}0.03 $\pm$ 0.12 $\pm$ 0.01 \\ 
$\overline{\Gamma}_{24}$ & $-0.10$ $\pm$ 0.10 $\pm$ 0.01 \\ 
$\overline{\Gamma}_{25}$ & \hphantom{$-$}0.03 $\pm$ 0.10 $\pm$ 0.01 \\ 
$\overline{\Gamma}_{26}$ & \hphantom{$-$}0.08 $\pm$ 0.11 $\pm$ 0.01 \\ 
$\overline{\Gamma}_{27}$ & \hphantom{$-$}0.14 $\pm$ 0.11 $\pm$ 0.01 \\ 
$\overline{\Gamma}_{28}$ & $-0.04$ $\pm$ 0.11 $\pm$ 0.01 \\ 
$\overline{\Gamma}_{29}$ & \hphantom{$-$}0.06 $\pm$ 0.15 $\pm$ 0.04 \\ 
$\overline{\Gamma}_{30}$ & $-0.21$ $\pm$ 0.15 $\pm$ 0.04 \\ 
$\overline{\Gamma}_{31}$ & $-0.07$ $\pm$ 0.16 $\pm$ 0.01 \\ 
$\overline{\Gamma}_{32}$ & $-0.16$ $\pm$ 0.17 $\pm$ 0.02 \\ 
$\overline{\Gamma}_{33}$ & $-0.04$ $\pm$ 0.17 $\pm$ 0.02 \\ 
$\overline{\Gamma}_{34}$ & \hphantom{$-$}0.15 $\pm$ 0.11 $\pm$ 0.01 \\ 
$\overline{\Gamma}_{35}$ & $-0.13$ $\pm$ 0.11 $\pm$ 0.01 \\ 
$\overline{\Gamma}_{36}$ & \hphantom{$-$}0.05 $\pm$ 0.11 $\pm$ 0.01 \\ 
$\overline{\Gamma}_{37}$ & \hphantom{$-$}0.05 $\pm$ 0.11 $\pm$ 0.01 \\ 
$\overline{\Gamma}_{38}$ & \hphantom{$-$}0.06 $\pm$ 0.11 $\pm$ 0.00 \\ 
$\overline{\Gamma}_{39}$ & $-0.08$ $\pm$ 0.11 $\pm$ 0.00 \\ 
$\overline{\Gamma}_{40}$ & \hphantom{$-$}0.15 $\pm$ 0.11 $\pm$ 0.01 \\ 
$\overline{\Gamma}_{41}$ & \hphantom{$-$}0.12 $\pm$ 0.11 $\pm$ 0.01 \\ 
\end{tabular}
\end{table}

\begin{figure}[!tb]
  \centering
  \includegraphics[width=0.32\textwidth]{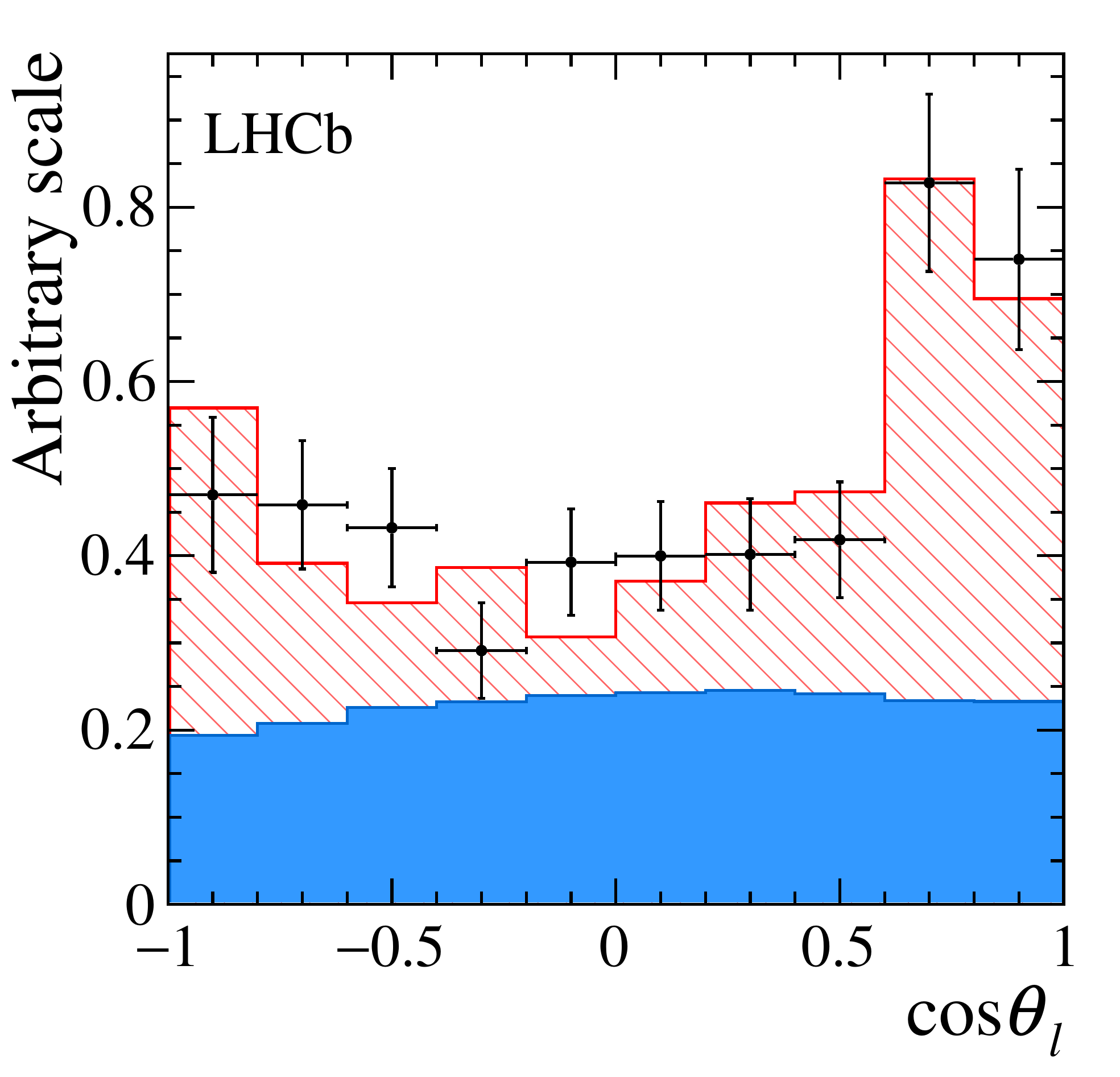}
  \includegraphics[width=0.32\textwidth]{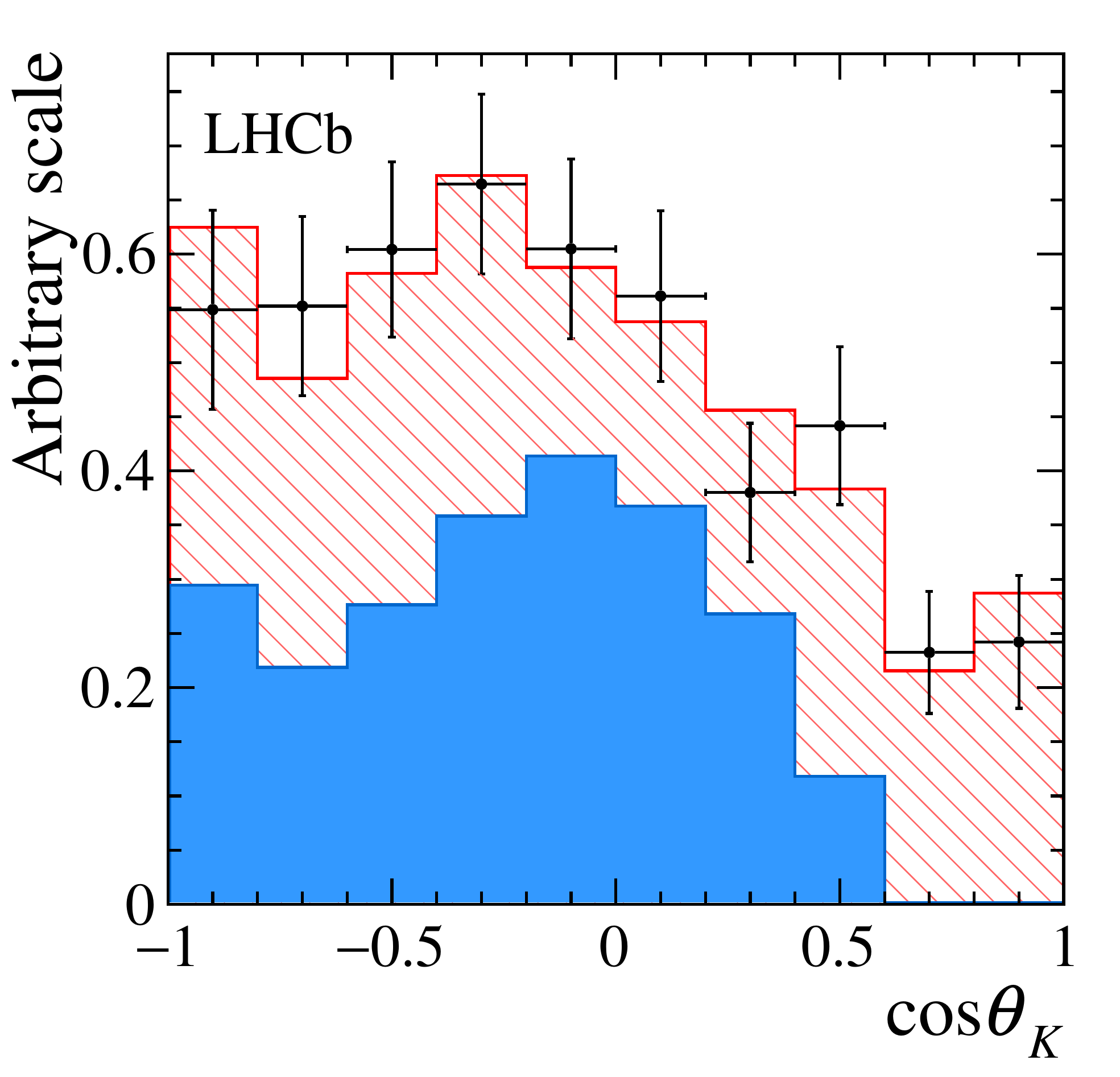}
  \includegraphics[width=0.32\textwidth]{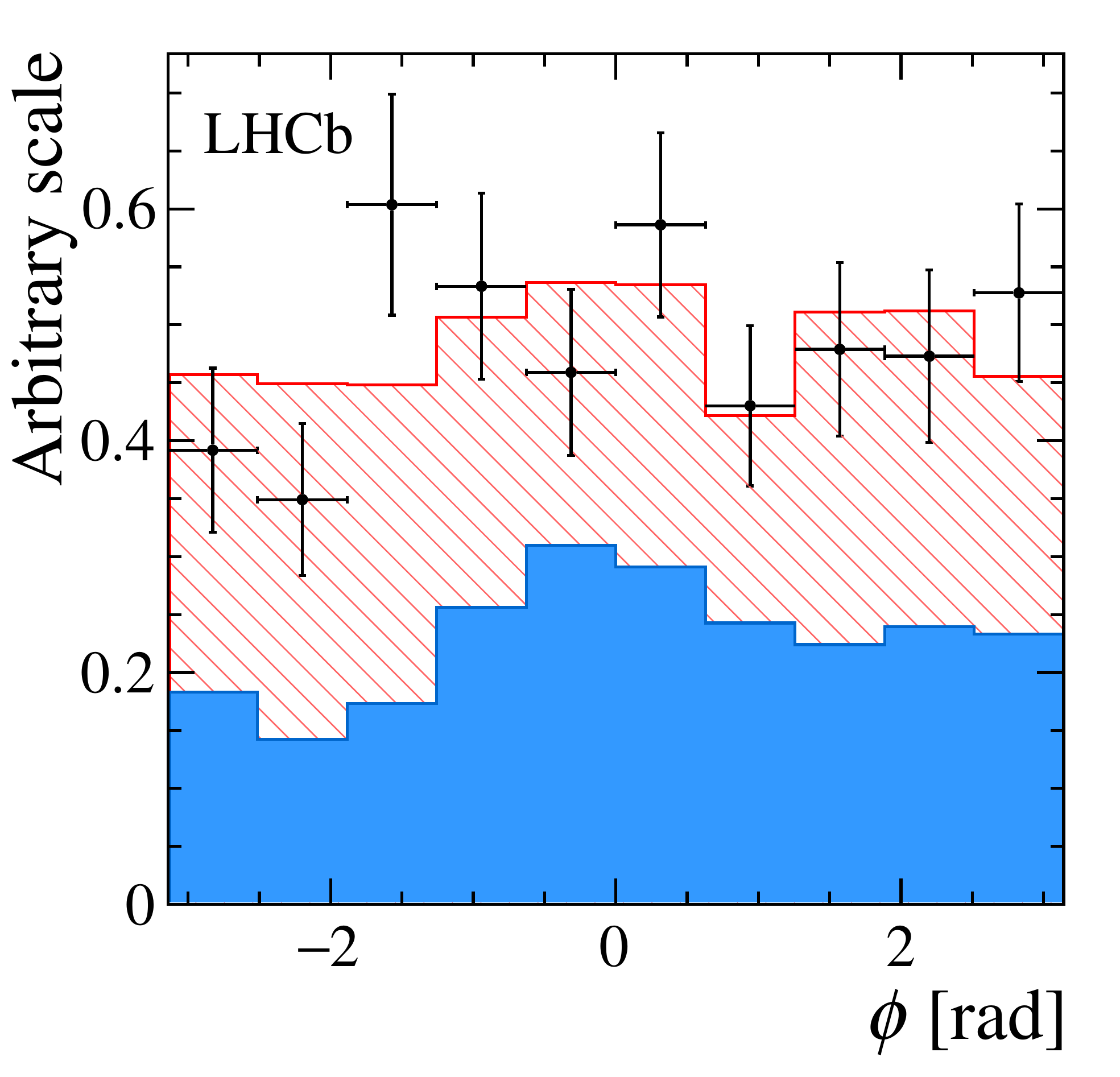}
  \caption{The distributions of each of the decay angles within the signal region. The acceptance-corrected data is represented by the points with error bars. The estimated signal distribution is shown by the blue, shaded histogram. The projected background from the upper mass sideband is shown by the red, hatched histogram, which is stacked onto the signal histogram.}
  \label{fig:gof_spd}
\end{figure}

The distributions of each of the decay angles within the signal region are shown in Fig.~\ref{fig:gof_spd}. The estimated signal distribution is derived 
from the moments model by evaluating the sum in Eq.~\ref{eqn:vector_moments}, which is found to provide a good representation of the data for each of the decay angles.

The D-wave fraction, $F_{\rm D}$, is estimated from the moments $\overline{\Gamma}_{5}$ and $\overline{\Gamma}_{10}$ as 

\begin{equation}
F_{\rm D} =  \displaystyle - \frac{7}{18} \left(2 \overline{\Gamma}_{5} + 5 \sqrt{5} \overline{\Gamma}_{10} \right).
\end{equation}

\noindent
Naively, one would expect a large D-wave contribution in this region, as was seen in the amplitude analysis of \BdToJPsiKpi~\cite{Chilikin:2014bkk}. However, in \BdToKpimm no significant D-wave contribution is seen and, with the limited statistics currently available, it is only possible to set an upper limit of $F_{\rm D}<0.29$ at 95\% confidence level using the approach in Ref.~\cite{Feldman:1997qc}. This might be an indication of a large breaking of QCD factorisation due to non-factorizable diagrams where additional gluons are exchanged between the $\Kp \pim$ and the \ccbar, before the \jpsi decays into $\mup \mun$. For electroweak penguins, similar effects could occur due to charm loops~\cite{Lyon:2014hpa}. Additionally, the values of the moments $\overline{\Gamma}_{2}$ and $\overline{\Gamma}_{3}$ imply the presence of large interference effects between the S- and P- or D-wave contributions.

\section{Systematic uncertainties}
\label{sec:systematics}

The main sources of systematic uncertainty for the measurements of the differential branching fraction and angular moments are described in detail below and summarised in Table~\ref{tab:systematics}. They are significantly smaller than the statistical uncertainties.

\begin{table}[!tb]
\caption{Summary of the main sources of systematic uncertainty for the differential branching fraction and the angular moments analysis. Typical ranges are quoted for the different \qsq bins used in the differential branching fraction measurement, and for the moments measured in the angular analysis.  The systematic uncertainties are significantly smaller than the statistical ones.}
\label{tab:systematics}
\begin{center}
\begin{tabular}{l|cc}
\multicolumn{1}{c|}{Source} & $d\BF/d\qsq \times 10^{-8}~[c^{4}/\gev^{2}]$ & $\overline{\Gamma}_{i}$ \\
\hline
Acceptance stat.\! uncertainty & 0.006--0.030 & 0.003--0.013 \\
Data-simulation differences & 0.001--0.014 & 0.001--0.007 \\
Peaking backgrounds & 0.013--0.026 & 0.001--0.040 \\
\hline
$\BF(\BdToJPsiKstP)$ & 0.033--0.110 & -- \\
\end{tabular}
\end{center}
\end{table} 

The differential branching fraction and angular moments analysis share several common systematic effects: the statistical uncertainty on the acceptance function due to the size of the simulated sample from which it is determined, differences between data and the simulated decays used to determine the acceptance function and contributions from residual peaking background candidates.  The differential branching fraction has, in addition, a systematic uncertainty due to the uncertainty on the branching fraction of the decay \BdToJPsiKstP, which is dominant and is shown separately in Table~\ref{tab:bf}. 

The size of the systematic uncertainties associated with the determination of the acceptance correction and residual peaking background contributions are evaluated using pseudoexperiments, in which samples are generated varying one or more parameters.  The differential branching fraction and each of the moments are evaluated using both the nominal model and the systematically varied models. 
In general, the systematic uncertainty is taken as the average of the difference between the nominal and varied models over a large number of pseudoexperiments. The exception to this is the statistical uncertainty of the acceptance function, due to the limited size of the simulated samples, for which the standard deviation is used instead.  For this, pseudoexperiments are generated where the acceptance is varied according to the covariance matrix of the moments of the acceptance function. 

The effect of differences between the data candidates and the simulated candidates is evaluated using pseudoexperiments, where candidates are generated with an acceptance determined from simulated candidates without applying the corrections for the differences between data and simulation described in Sec.~\ref{sec:Detector}.

The effect of residual peaking background contributions is evaluated using pseudoexperiments, where peaking background components are generated in addition to the signal and the combinatorial background.  The angular distributions of the peaking backgrounds are taken from data by isolating the decays using dedicated selections.

All other sources of systematic uncertainties investigated, such as the choice of the \mkpimm signal model and the resolution in the angular variables, are found to have a negligible impact.

\section{Conclusions}
\label{sec:conclusions}

This paper presents measurements of the differential branching fraction and angular moments of the decay $\decay{\B^0}{K^{+}\pi^{-}\mu^{+}\mu^{-}}$ in the $K^{+}\pi^{-}$ invariant mass range \mbox{$1330<m(\Kp\pim)<1530~\mathrm{\,Me\kern -0.1em V\!/}c^{2}$}.  The data sample corresponds to an integrated luminosity of 3$\ensuremath{\mbox{\,fb}^{-1}}\xspace$ of $pp$ collision data collected by the LHCb experiment.  The differential branching fraction is reported in five narrow \qsq bins between 0.1 and 8.0\gevgevcccc and in the range $1.1<\qsq<6.0\gevgevcccc$, where an angular moments analysis is also performed.

The measured values of the angular observables $\overline{\Gamma}_{2}$ and $\overline{\Gamma}_{3}$ point towards the presence of large interference effects between the S- and P- or D-wave contributions. Using only $\overline{\Gamma}_{5}$ and $\overline{\Gamma}_{10}$ it is possible to estimate the D-wave fraction, $F_{\rm D}$,  yielding an upper limit of $F_{\rm D}<0.29$ at 95\% confidence level. This value is lower than naively expected from amplitude analyses of \BdToJPsiKpi decays~\cite{Chilikin:2014bkk}.

The underlying Wilson coefficients may be extracted from the normalised moments and covariance matrix presented in this analysis, when combined with a prediction for the form factors. While first estimates for the form factors are given in Ref.~\cite{Lu:2011jm}, no interpretation of the results in terms of the Wilson coefficients is made at this time. With additional input from theory, these results could provide further contributions to understanding the pattern of deviations with respect to SM predictions that has been observed in other $b\to s\mu\mu$ transitions.

% Do not include this in analysis note and conference reports
\section*{Acknowledgements}

% The text below are the acknowledgements as approved by the collaboration
% board. Extending the acknowledgements to include individuals from outside the
% collaboration who have contributed to the analysis should be approved by the
% EB. The extra acknowledgements are normally placed before the standard 
% acknowledgements, unless it matches better with the text of the standard 
% acknowledgements to put them elsewhere. They should be included in the draft 
% for the first circulation. Except in exceptional circumstances, to be approved by the
% EB chair, authors of the paper should not be named in extended acknowledgements.
 
\noindent We express our gratitude to our colleagues in the CERN
accelerator departments for the excellent performance of the LHC. We
thank the technical and administrative staff at the LHCb
institutes. We acknowledge support from CERN and from the national
agencies: CAPES, CNPq, FAPERJ and FINEP (Brazil); NSFC (China);
CNRS/IN2P3 (France); BMBF, DFG and MPG (Germany); INFN (Italy); 
FOM and NWO (The Netherlands); MNiSW and NCN (Poland); MEN/IFA (Romania); 
MinES and FASO (Russia); MinECo (Spain); SNSF and SER (Switzerland); 
NASU (Ukraine); STFC (United Kingdom); NSF (USA).
We acknowledge the computing resources that are provided by CERN, IN2P3 (France), KIT and DESY (Germany), INFN (Italy), SURF (The Netherlands), PIC (Spain), GridPP (United Kingdom), RRCKI and Yandex LLC (Russia), CSCS (Switzerland), IFIN-HH (Romania), CBPF (Brazil), PL-GRID (Poland) and OSC (USA). We are indebted to the communities behind the multiple open 
source software packages on which we depend.
Individual groups or members have received support from AvH Foundation (Germany),
EPLANET, Marie Sk\l{}odowska-Curie Actions and ERC (European Union), 
Conseil G\'{e}n\'{e}ral de Haute-Savoie, Labex ENIGMASS and OCEVU, 
R\'{e}gion Auvergne (France), RFBR and Yandex LLC (Russia), GVA, XuntaGal and GENCAT (Spain), Herchel Smith Fund, The Royal Society, Royal Commission for the Exhibition of 1851 and the Leverhulme Trust (United Kingdom).

\clearpage

{\noindent\normalfont\bfseries\Large Appendices}

\appendix

\section{Angular distribution}
\label{sec:appendix:angular-distribution}

The transversity-basis moments of the 41 orthonormal angular functions defined in Eq.~\ref{eqn:vector_moments}
 are shown in Table~\ref{table:spd_mom_trans}. The orthonormal angular basis is constructed out of spherical harmonics, \mbox{$Y^m_l \equiv Y^m_l (\thetal,\phi)$}, and reduced spherical harmonics, \mbox{${P^m_l \equiv \sqrt{2 \pi}Y^m_l(\thetak,0)}$}. The S-, P- and D-wave transversity amplitudes are denoted as $S^{\{L,R\}}$, $H^{\{L,R\}}_{\{0,\parallel,\perp\}}$ and $D^{\{L,R\}}_{\{0,\parallel,\perp\}}$, respectively.

It should be noted that in addition to dependence on the amplitudes there is an overall kinematic factor of ${\bf k}\qsq$, where ${\bf k}$ is the $\Bz$ break-up momentum given by
\begin{align}
{\bf k} = \displaystyle \sqrt{ \frac{\left(m^2_B-q^2 +m^2(\Kp\pim) \right)^2}{4m^2_B} - m^2(\Kp\pim)  },
\end{align}
and $m_B$ is the $\Bz$ mass.

\begin{table}[!tb]
\caption{The transversity-basis moments of the 41 orthonormal angular functions $f_i(\Omega)$ in Eq.~\ref{eqn:vector_moments}~\cite{spd-paper}. The amplitudes correspond to the $\Bzb$ decay.}
\centering
\resizebox{0.95\textwidth}{!}{
\begin{tabular}{c|c|c|c} 
 $i$    &   $f_i(\Omega)$             & $\Gamma^{L, {\rm tr}}_i(\qsq)/{\bf k}\qsq $ & $\eta^{L\to R}_i$  \\ \hline \hline
 1   &   $P^0_0 Y^0_0$     &  $\left[ \hzsq + \hpasq + \hpesq + \ssq + \dzsq + \dpasq + \dpesq\right]$ & $+1$ \\ \hline 
 2   &   $P^0_1 Y^0_0$     &  $2\left[\frac{2}{\sqrt{5}} \rhzdz + \rshz + \sqrt{\frac{3}{5}}  \rel( H^L_\parallel D^{L\ast}_\parallel + H^L_\perp D^{L\ast}_\perp  )\right]$ & $+1$ \\ \hline 
 3   &   $P^0_2 Y^0_0$     &  $\frac{\sqrt{5}}{7}$ (\dpasq + \dpesq) - $\frac{1}{\sqrt{5}}$ (\hpasq + \hpesq) + $\frac{2}{\sqrt{5}}$ \hzsq  + $\frac{10}{7\sqrt{5}}$ \dzsq + $2$ \rsdz & $+1$ \\  \hline
 4   &   $P^0_3 Y^0_0$     &  $\frac{6}{\sqrt{35}} \left[ - \rel(H^L_\parallel D^{L\ast}_\parallel +  H^L_\perp D^{L\ast}_\perp)  + \sqrt{3} \rhzdz  \right]$ & $+1$\\  \hline
 5   &   $P^0_4 Y^0_0$     &  $\frac{2}{7} \left[ -2 (\dpasq + \dpesq) + 3 \dzsq \right] $ & $+1$\\  \hline
 6   &   $P^0_0 Y^0_2$     &  $\frac{1}{2 \sqrt{5}} \left[ (\dpasq + \dpesq) + (\hpasq + \hpesq) - 2 \ssq - 2 \dzsq - 2 \hzsq \right]$ & $+1$ \\  \hline
 7   &   $P^0_1 Y^0_2$     &  $\left[ \frac{\sqrt{3}}{5} \rel(H^L_\parallel D^{L\ast}_\parallel  + H^L_\perp D^{L\ast}_\perp) - \frac{2}{\sqrt{5}} \rel(S^L H^{L\ast}_0)  - \frac{4}{5} \rel(H^L_0 D^{L\ast}_0)\right] $ & $+1$  \\ \hline
 8   &   $P^0_2 Y^0_2$     &  $ \left[ \frac{1}{14} (\dpasq + \dpesq) - \frac{2}{7} \dzsq - \frac{1}{10} (\hpasq + \hpesq) - \frac{2}{5} \hzsq - \frac{2}{\sqrt{5}} \rsdz \right]$ & $+1$  \\  \hline
 9   &   $P^0_3 Y^0_2$     &  $ - \frac{3}{5 \sqrt{7}} \left[ \rel( H^L_\parallel D^{L \ast}_\parallel + H^L_\perp D^{L \ast}_\perp) + 2 \sqrt{3} \rel(H^L_0 D^{L \ast}_0 ) \right] $ & $+1$\\  \hline
 10  &   $P^0_4 Y^0_2$     &  $ -\frac{2}{7 \sqrt{5}}  \left[ \dpasq + \dpesq + 3 \dzsq \right] $ & $+1$  \\  \hline
 11  &   $P^1_1 \sqrt{2}\rel(Y^1_2)$ &  $-\frac{3}{\sqrt{10}} \left[ \sqrt{\frac{2}{3}} \rel(H^L_\parallel S^{L \ast}) - \sqrt{\frac{2}{15}} \rel(H^L_\parallel D^{L \ast}_0  ) + \sqrt{\frac{2}{5}} \rel(D^L_\parallel H^{L \ast}_0 ) \right] $  & $+1$\\  \hline
 12  &   $P^1_2 \sqrt{2}\rel(Y^1_2)$ &  $-\frac{3}{5} \left[ \rel( H^L_\parallel H^{L \ast}_0)  + \sqrt{\frac{5}{3}} \rel (D^L_\parallel S^{L \ast})  + \frac{5}{7 \sqrt{3}} \rel(D^L_\parallel D^{L\ast}_0) \right] $ & $+1$ \\  \hline
 13  &   $P^1_3 \sqrt{2}\rel(Y^1_2)$ &  $-\frac{6}{5 \sqrt{14}} \left[2 \rel(D^L_\parallel H^{L\ast}_0)  + \sqrt{3} \rel(H^L_\parallel D^{L\ast}_0 ) \right] $ & $+1$ \\  \hline
 14  &   $P^1_4 \sqrt{2}\rel(Y^1_2)$ &  $- \frac{6}{7\sqrt{2}} \rel(D^L_\parallel D^{L\ast}_0)$  & $+1$ \\  \hline
 15  &   $P^1_1 \sqrt{2}\img(Y^1_2)$ &  $3 \left[ \frac{1}{\sqrt{15}} \img(H^L_\perp S^{L\ast}) + \frac{1}{5} \img(D^L_\perp H^{L\ast}_0) - \frac{1}{5 \sqrt{3}}  \img(H^L_\perp D^{L\ast}_0) \right]  $  & $+1$ \\  \hline
 16  &   $P^1_2 \sqrt{2}\img(Y^1_2)$ &  $ 3\left[ \frac{1}{7 \sqrt{3}} \img(D^L_\perp D^{L\ast}_0)  + \frac{1}{5} \img(H^L_\perp H^{L\ast}_0)  + \frac{1}{\sqrt{15}} \img(D^L_\perp S^{L\ast})   \right] $  & $+1$ \\  \hline
 17  &   $P^1_3 \sqrt{2}\img(Y^1_2)$ &  $\frac{6}{5 \sqrt{14}} \left[ 2 \img(D^L_\perp H^{L\ast}_0)  + \sqrt{3} \img(H^L_\perp D^{L\ast}_0) \right]  $   & $+1$ \\  \hline
 18  &   $P^1_4 \sqrt{2}\img(Y^1_2)$ &  $\frac{6}{7\sqrt{2}} \img(D^L_\perp D^{L\ast}_0)$  & $+1$ \\  \hline
 19  &   $P^0_0 \sqrt{2}\rel(Y^2_2)$ &  $-\frac{3}{2\sqrt{15}} \left[ (\hpasq - \hpesq) + (\dpasq - \dpesq) \right] $  & $+1$ \\  \hline
 20  &   $P^0_1 \sqrt{2}\rel(Y^2_2)$ &  $-\frac{3}{5} \left[ \rel(H^L_\parallel D^{L\ast}_\parallel)   - \rel(D^L_\perp H^{L\ast}_\perp) \right] $  & $+1$ \\  \hline
 21  &   $P^0_2 \sqrt{2}\rel(Y^2_2)$ &  $\frac{\sqrt{3}}{2} \left[ - \frac{1}{7} (\dpasq - \dpesq)   + \frac{1}{5} ( \hpasq - \hpesq ) \right] $  & $+1$ \\  \hline
 22  &   $P^0_3 \sqrt{2}\rel(Y^2_2)$ &  $\frac{3}{5} \sqrt{ \frac{3}{7}} \left[ \rel(H^L_\parallel D^{L\ast}_\parallel)   - \rel(D^L_\perp H^{L\ast}_\perp) \right] $  & $+1$ \\  \hline
 23  &   $P^0_4 \sqrt{2}\rel(Y^2_2)$ &  $\frac{2}{7} \sqrt{ \frac{3}{5}}  (\dpasq - \dpesq) $ & $+1$ \\  \hline
 24  &   $P^0_0 \sqrt{2}\img(Y^2_2)$ &  $\sqrt{\frac{3}{5}} \left[ \img(H^L_\perp H^{L\ast}_\parallel) + \img(D^L_\perp D^{L\ast}_\parallel) \right] $   & $+1$ \\  \hline
 25  &   $P^0_1 \sqrt{2}\img(Y^2_2)$ &  $\frac{3}{5} \img(  H^L_\perp D^{L\ast}_\parallel +  D^L_\perp H^{L\ast}_\parallel )  $  & $+1$ \\ \hline
 26  &   $P^0_2 \sqrt{2}\img(Y^2_2)$ &  $ \sqrt{3} \left[\frac{1}{7} \img(D^L_\perp D^{L\ast}_\parallel)   - \frac{1}{5} \img(H^L_\perp H^{L\ast}_\parallel)\right] $  & $+1$ \\ \hline
 27  &   $P^0_3 \sqrt{2}\img(Y^2_2)$ &  $-\frac{3}{5} \sqrt{ \frac{3}{7}}  \img(D^L_\perp H^{L\ast}_\parallel + H^L_\perp D^{L\ast}_\parallel)  $  & $+1$ \\ \hline
 28  &   $P^0_4 \sqrt{2}\img(Y^2_2)$ &  $-\frac{4}{7} \sqrt{\frac{3}{5}}  \img(D^L_\perp D^{L\ast}_\parallel) $   & $+1$ \\ \hline \hline
 29  &   $P^0_0 Y^0_1$     &  $-\sqrt{3}\left[ \rel(H^L_\perp H^{L\ast}_\parallel) + \rel(D^L_\perp D^{L\ast}_\parallel) \right]$  & $-1$ \\ \hline
 30  &   $P^0_1 Y^0_1$     &  $-\frac{3}{\sqrt{5}} \rel( H^L_\perp D^{L\ast}_\parallel + H^L_\parallel D^{L\ast}_\perp ) $ & $-1$ \\ \hline
 31  &   $P^0_2 Y^0_1$     &  $-\frac{3}{\sqrt{15}} \left[ \frac{5}{7} \rel(D^L_\perp D^{L\ast}_\parallel) - \rel(H^L_\perp H^{L\ast}_\parallel)  \right]$ & $-1$ \\ \hline
 32  &   $P^0_3 Y^0_1$     &  $\frac{9}{\sqrt{105}}  \rel(H^L_\perp D^{L\ast}_\parallel  + H^L_\parallel D^{L\ast}_\perp ) $ & $-1$ \\ \hline
 33  &   $P^0_4 Y^0_1$     &  $\frac{4\sqrt{3}}{7} \rel(D^L_\perp D^{L\ast}_\parallel)$  & $-1$ \\ \hline
 34  &   $P^1_1 \sqrt{2}\rel(Y^1_1)$   & $\sqrt{\frac{3}{5}} \left[ \sqrt{5} \rel(H^L_\perp S^{L \ast})  + \sqrt{3} \rel(D^L_\perp H^{L\ast}_0)  - \rel(H^L_\perp D^{L \ast}_0) \right]$  & $-1$ \\ \hline
 35  &   $P^1_2 \sqrt{2}\rel(Y^1_1)$   & $ 3 \left[ \frac{1}{\sqrt{5}} \rel(H^L_\perp H^{L \ast}_0)  + \frac{1}{\sqrt{3}} \rel(D^L_\perp S^{L\ast})  + \frac{5}{21} \sqrt{\frac{3}{5}} \rel(D^L_\perp D^{L \ast}_0 ) \right] $  & $-1$ \\ \hline
 36  &   $P^1_3 \sqrt{2}\rel(Y^1_1)$   & $ \frac{6}{\sqrt{70}} \left[ 2 \rel(D^L_\perp H^{L \ast}_0)  + \sqrt{3} \rel(H^L_\perp D^{L\ast}_0) \right]$  & $-1$ \\ \hline
 37  &   $P^1_4 \sqrt{2}\rel(Y^1_1)$   & $\frac{3 \sqrt{10}}{7} \rel(D^L_\perp D^{L \ast}_0 ) $  & $-1$ \\ \hline
 38  &   $P^1_1 \sqrt{2}\img(Y^1_1)$   & $-\sqrt{\frac{3}{5}} \left[ \sqrt{5} \img ( H^L_\parallel S^{L\ast}) + \sqrt{3} \img(D^L_\parallel H^{L \ast}_0) - \img(H^L_\parallel D^{L \ast}_0)  \right]  $  & $-1$ \\ \hline
 39  &   $P^1_2 \sqrt{2}\img(Y^1_1)$   & $ -\sqrt{\frac{3}{5}} \left[ \sqrt{3} \img(H^L_\parallel H^{L \ast}_0)  + \sqrt{5} \img(D^L_\parallel S^{L\ast})  + \frac{5}{7} \img(D^L_\parallel D^{L \ast}_0 )\right] $  & $-1$ \\ \hline
 40  &   $P^1_3 \sqrt{2}\img(Y^1_1)$   & $ -6\sqrt{\frac{1}{70}}\left[ 2\img(D^L_\parallel H^{L \ast}_0)  + \sqrt{3} \img(H^L_\parallel D^{L\ast}_0)\right]$   & $-1$ \\ \hline
 41  &   $P^1_4 \sqrt{2}\img(Y^1_1)$   & $-\frac{3}{7} \sqrt{10} \img(D^L_\parallel D^{L\ast}_0) $  & $-1$ \\ \hline
\end{tabular}

}
\label{table:spd_mom_trans}
\end{table}

\clearpage

\section{Mass distributions}
\label{sec:appendix:massfit}

\begin{figure}[!h]
 \centering
 \includegraphics[width=0.48\linewidth]{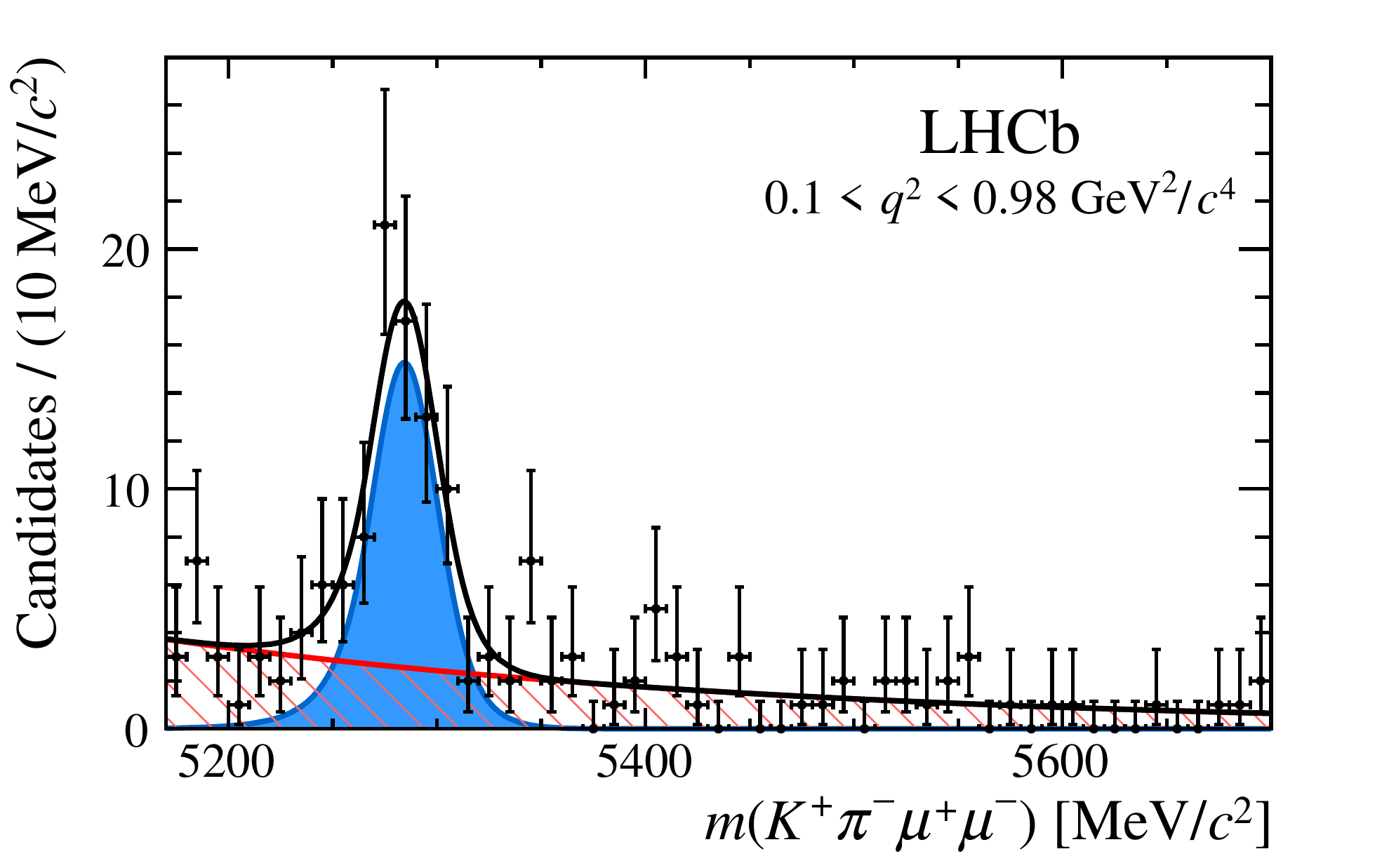}
 \includegraphics[width=0.48\linewidth]{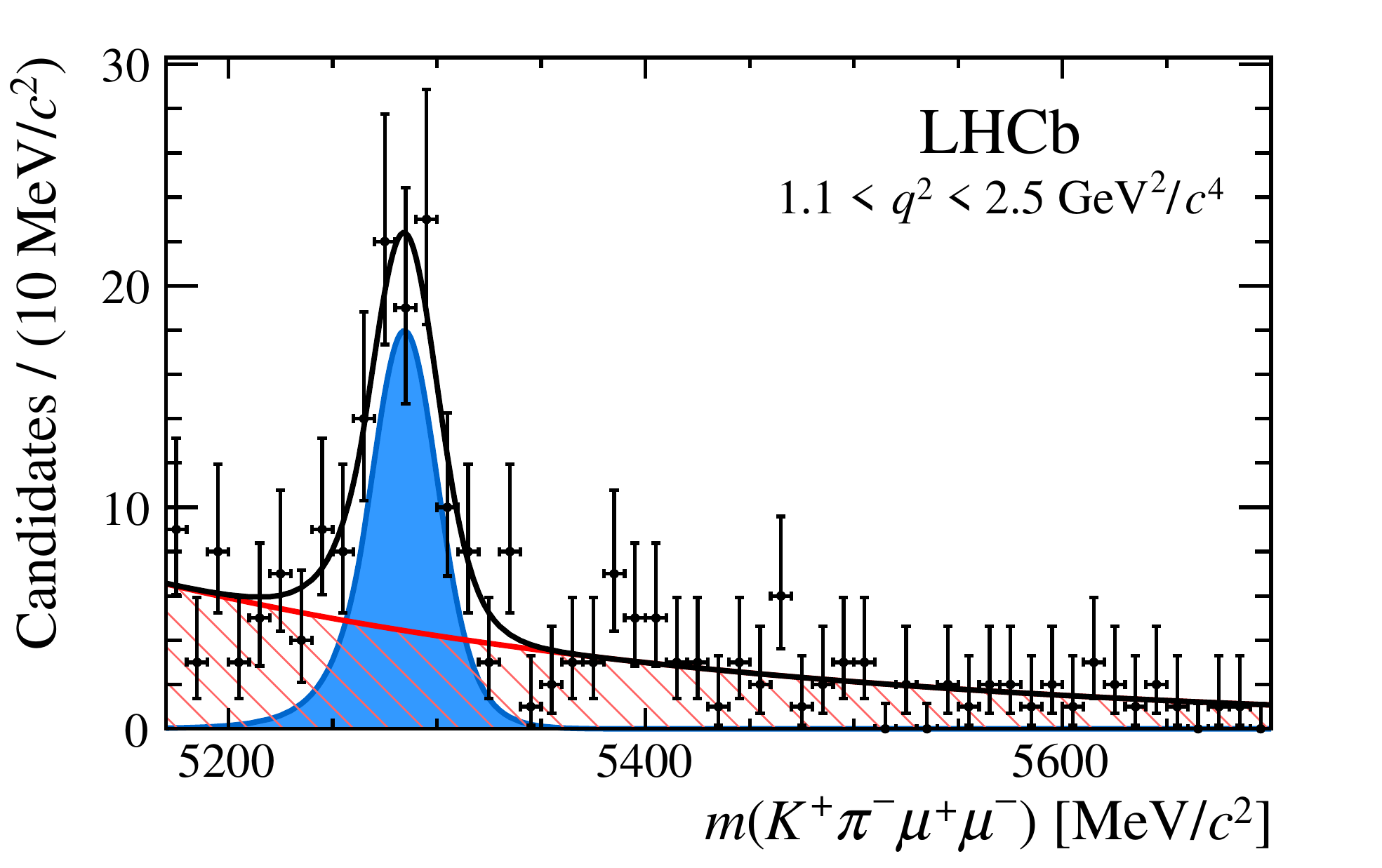}
 \includegraphics[width=0.48\linewidth]{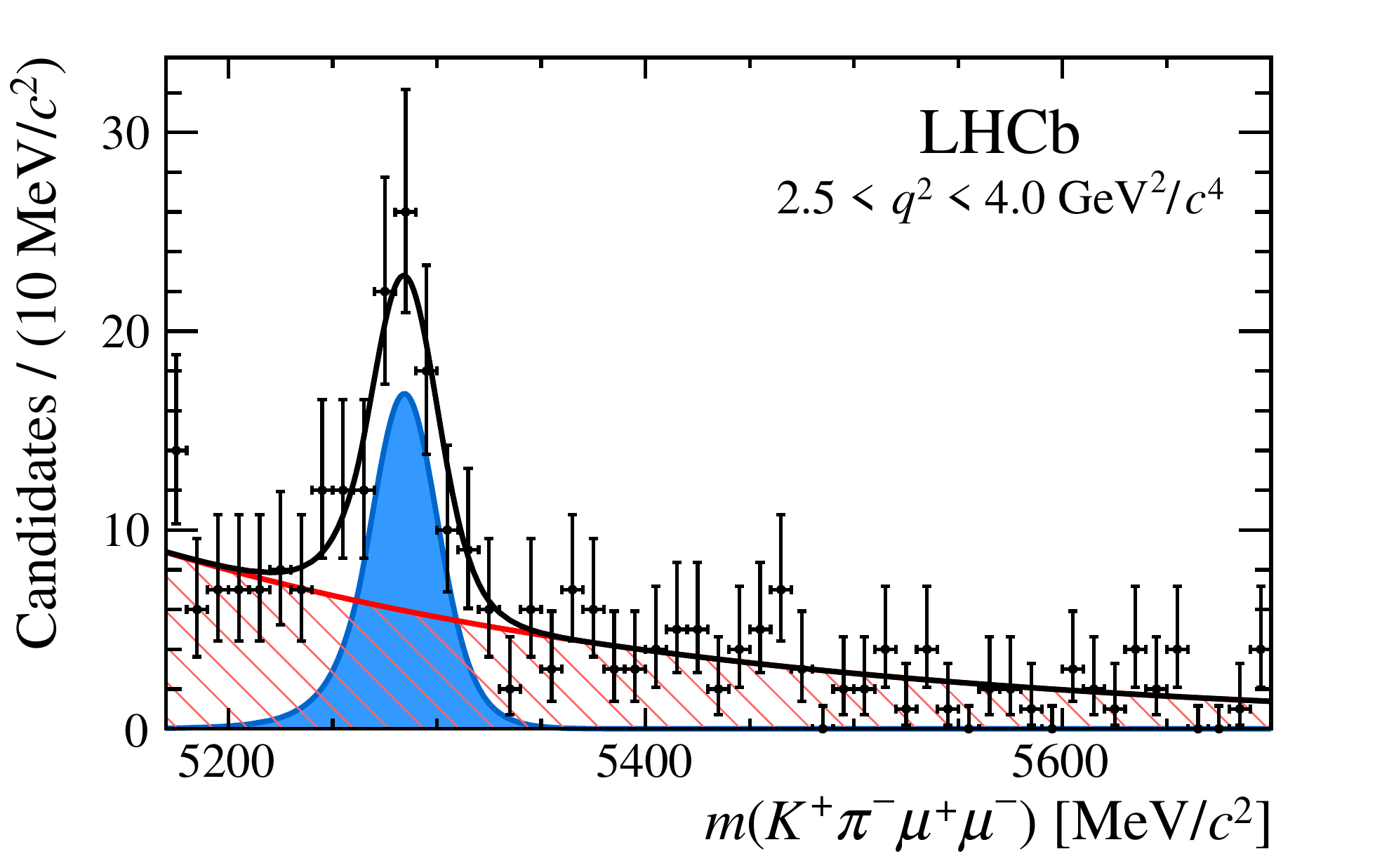}
 \includegraphics[width=0.48\linewidth]{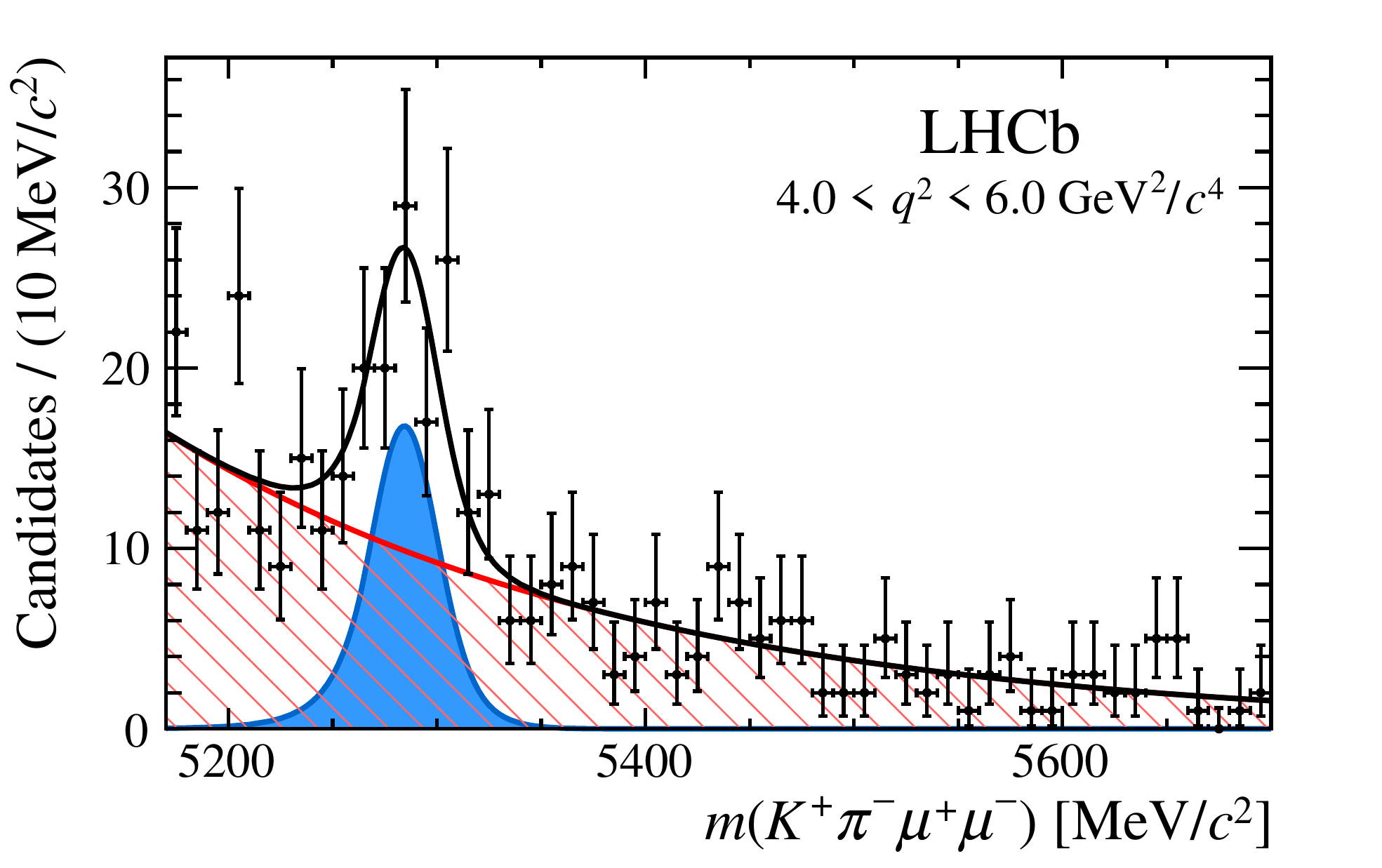}
 \includegraphics[width=0.48\linewidth]{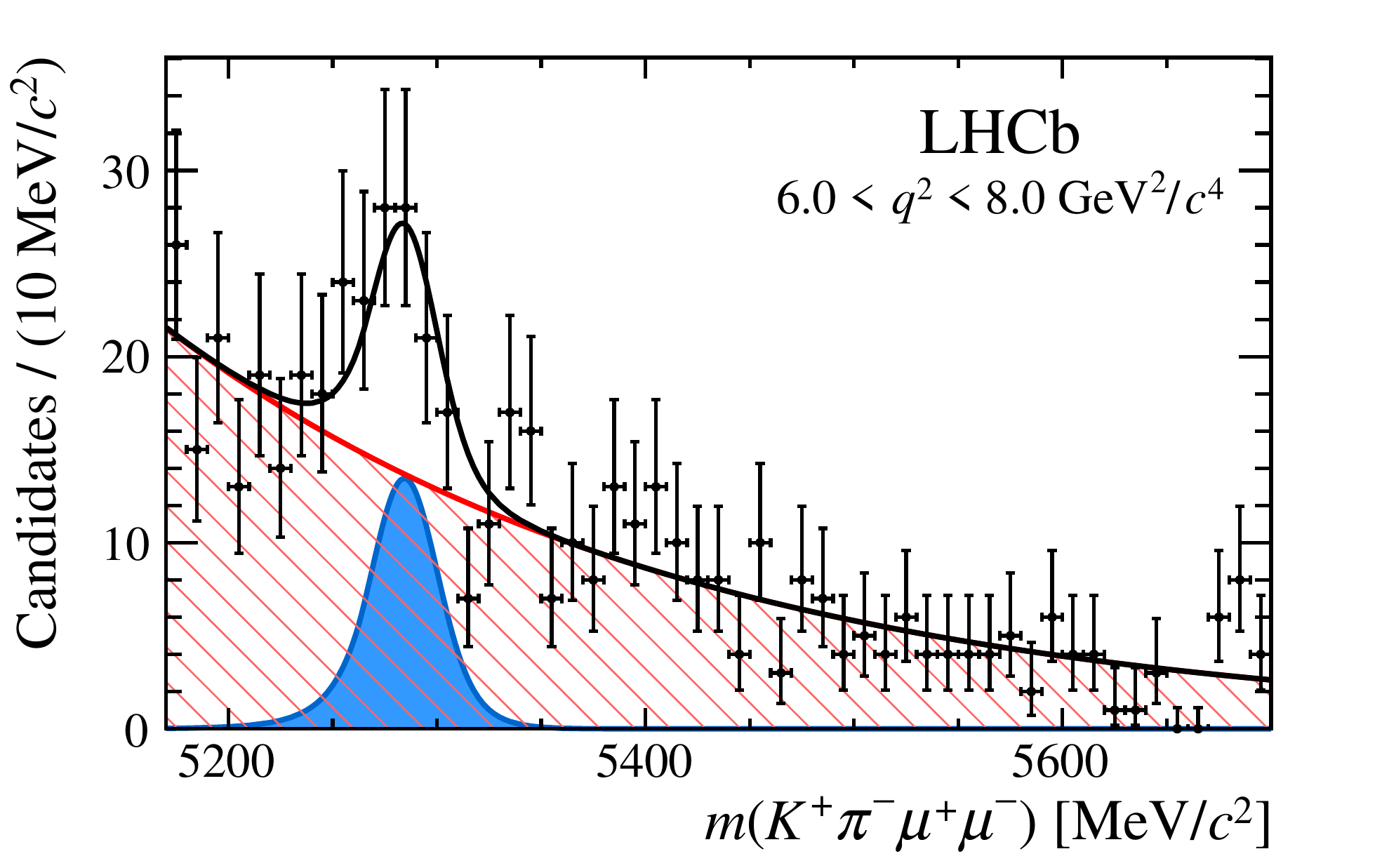}
 \includegraphics[width=0.48\linewidth]{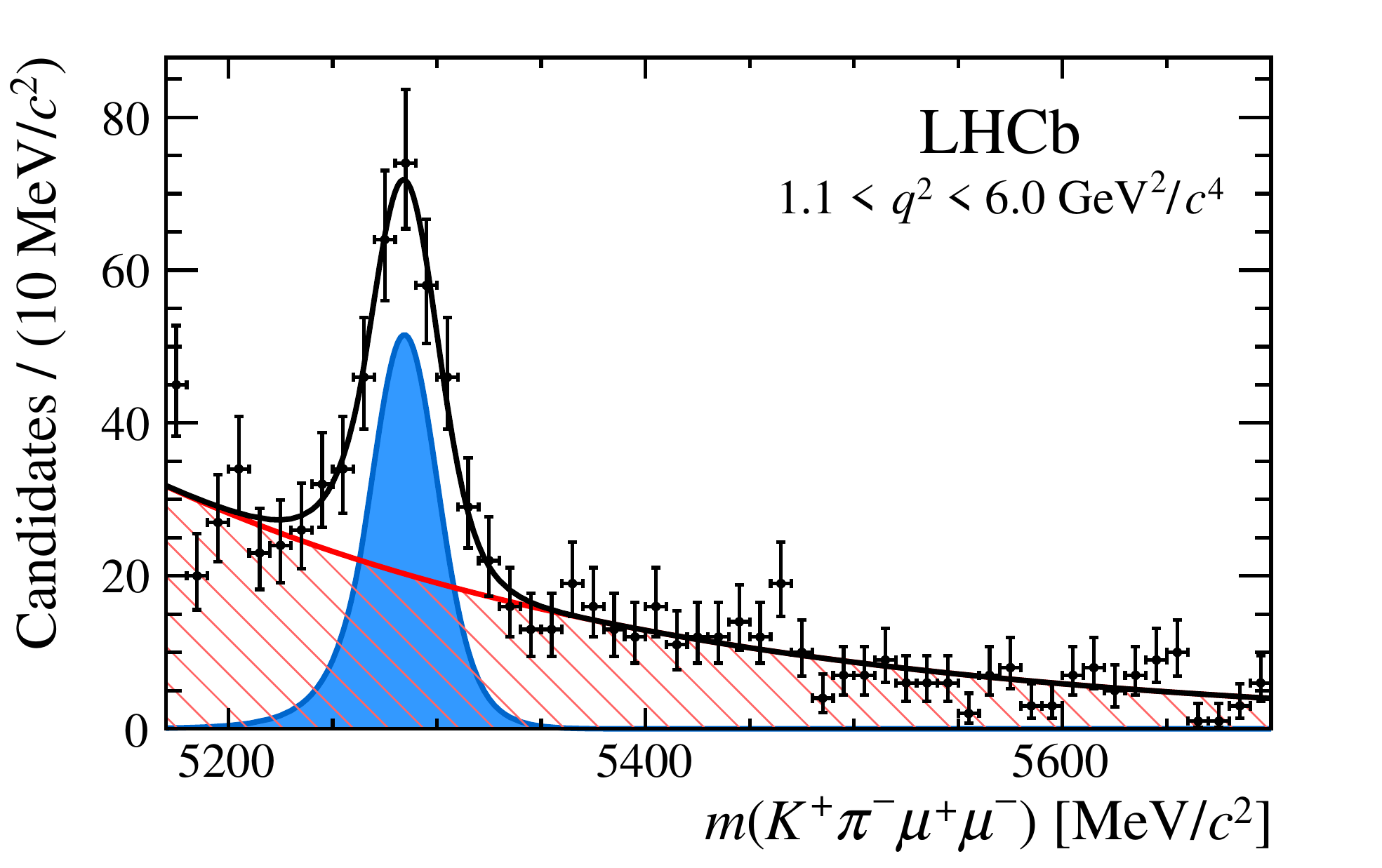}

 \caption{Invariant mass \mkpimm distributions of the signal decay \BdToKpimm in  each of the \qsq bins used for the differential branching fraction measurement. The solid black line represents the total fitted function.  The individual components of the signal (blue, shaded area) and combinatorial background (red, hatched area) are also shown.}
\label{fig:massfit:bins}
\end{figure}

Figure~\ref{fig:massfit:bins} shows the fits to the \mkpimm distribution in each of the \qsq bins used for the differential branching fraction measurement.

% This should be taken out in the final paper
%\input{supplementary-app}
\clearpage
\addcontentsline{toc}{section}{References}
\setboolean{inbibliography}{true}
\bibliographystyle{LHCb}
\bibliography{main,LHCb-PAPER,LHCb-CONF,LHCb-DP,LHCb-TDR}

\newpage

% Author List ----------------------------                                                                                                                                                                                                                                                                                                
%  You need to get a new author list!                                                                                                                                                                                                                                                                                                    

%\input{LHCb_HD_authorlist_2014-06-20}
 
 \newpage
 \centerline{\large\bf LHCb collaboration}
\begin{flushleft}
\small
R.~Aaij$^{40}$,
B.~Adeva$^{39}$,
M.~Adinolfi$^{48}$,
Z.~Ajaltouni$^{5}$,
S.~Akar$^{6}$,
J.~Albrecht$^{10}$,
F.~Alessio$^{40}$,
M.~Alexander$^{53}$,
S.~Ali$^{43}$,
G.~Alkhazov$^{31}$
% ...

{\footnotesize \it
% ...
$ ^{5}$Clermont Universit{\'e}, Universit{\'e} Blaise Pascal, CNRS/IN2P3, LPC, Clermont-Ferrand, France\\
$ ^{6}$CPPM, Aix-Marseille Universit{\'e}, CNRS/IN2P3, Marseille, France\\
% ...
$ ^{10}$Fakult{\"a}t Physik, Technische Universit{\"a}t Dortmund, Dortmund, Germany\\
% ...
$ ^{31}$Petersburg Nuclear Physics Institute (PNPI), Gatchina, Russia\\
% ...
$ ^{39}$Universidad de Santiago de Compostela, Santiago de Compostela, Spain\\
$ ^{40}$European Organization for Nuclear Research (CERN), Geneva, Switzerland\\
% ...
$ ^{43}$Nikhef National Institute for Subatomic Physics, Amsterdam, The Netherlands\\
% ...
$ ^{48}$H.H. Wills Physics Laboratory, University of Bristol, Bristol, United Kingdom\\
% ...
$ ^{53}$School of Physics and Astronomy, University of Glasgow, Glasgow, United Kingdom\\
% ...
}
\end{flushleft}

% The author list for journal publications is generated from the Membership Database shortly after 'approval to go to paper' has been given.
% It will be sent to you by email shortly after a paper number has been assigned.
% The author list should be included already at first circulation,
% to allow new members of the collaboration to verify that they have been included correctly.
% Occasionally a misspelled name is corrected, or associated institutions become full members.
% Therefore an updated author list will be sent to you after the final EB review of the paper.
% In case line numbering doesn't work well after including the authorlist, try moving the \verb!\bigskip! after the last author to a separate line.

% The authorship for Conference Reports should be ``The LHCb                                                                                                                                                                                                                                                                                
%   collaboration'', with a footnote giving the name(s) of the contact
%   author(s), but without the full list of collaboration names.

\end{document}